\documentclass[a4paper,12pt]{article}
\usepackage{authblk}
\usepackage[dvipsnames]{xcolor}
\usepackage{latexsym,amsmath,amssymb,amsfonts,amsthm,dsfont,enumerate,natbib,bbm,threeparttable,rotating, pdflscape,verbatim,amssymb}
\usepackage{algorithm,algcompatible,algpseudocode,booktabs,xcolor,tikz}
\usetikzlibrary{arrows,shapes,chains}
\usepackage[normalem]{ulem}

\usepackage[colorlinks,linkcolor=red,citecolor=blue,urlcolor=blue,breaklinks]{hyperref}
\usepackage{cleveref}

\newcommand{\norm}[1]{\|#1\|}
\newcommand{\rank}[1]{\text{rank}(#1)}
\newcommand{\abs}[1]{\vert #1 \vert}
\newcommand{\inner}[1]{\langle#1\rangle}
\def\code#1{\texttt{#1}}

\newcommand{\bDelta}{\mbox{\boldmath $\Delta$}}
\newcommand{\bdelta}{\mbox{\boldmath $\delta$}}
\newcommand{\bSigma}{\mbox{\boldmath $\Sigma$}}
\newcommand{\bGamma}{\mbox{\boldmath $\Gamma$}}

\algnewcommand\algorithmicpara{\textbf{estimate in parallel for}}
\algblockdefx[PARA]{Para}{EndPara}[1]{\algorithmicpara\ #1}{\algorithmicend}
\algnewcommand\algorithmicdivide{\textbf{divide}}
\algnewcommand\algorithmicextract{\textbf{matrix decomposition}}
\algblockdefx[DIVIDE]{Divide}{EndDivide}{\algorithmicdivide}{\algorithmicend}
\algblockdefx[EXTRACT]{Extraction}{EndExtract}{\algorithmicextract}{\algorithmicend}

\algnewcommand{\algorithmicendif}{\textbf{end}}
\algblockdefx[IF]{If}{EndIf}[1]{\algorithmicif\ #1\ \algorithmicthen}{\algorithmicendif}

\newtheorem{theorem}{Theorem}[section]
\newtheorem{lemma}{Lemma}%[section]
\newtheorem{condition}{Condition}[section]

\newtheorem{proposition}[theorem]{Proposition}

\def\trans{^{\rm T}}
\def\strans{^{*\rm T}}

\def\var{\text{Var}}

\def\M{{\mathbb M}}
\def\R{{\mathbb R}}
\def\T{{\mathbb T}}

\def\bT{{\bf T}}
\def\C{{\bf C}}
\def\X{{\bf X}}
\def\Y{{\bf Y}}
\def\U{{\bf U}}
\def\V{{\bf V}}
\def\D{{\bf D}}
\def\I{{\bf I}}
\def\proj{{\bf P}}
\def\E{{\bf E}}
\def\A{{\bf A}}
\def\B{{\bf B}}
\def\W{{\bf W}}
\def\L{{\bf L}}
\def\S{{\bf S}}
\def\Z{{\bf Z}}
\def\Q{{\bf Q}}

\def\x{{\bf x}}
\def\y{{\bf y}}
\def\e{{\bf e}}
\def\c{{\bf c}}
\def\u{{\bf u}}
\def\v{{\bf v}}
\def\z{{\bf z}}
\def\a{{\bf a}}
\def\b{{\bf b}}

\def\0{{\bf 0}}

\RequirePackage{lineno}
\providecommand{\keywords}[1]{\textbf{\textit{Keywords---}} #1}
  
\begin{document}

%\title{Parallel integrative learning for large-scale multi-response regression with incomplete outcomes}
%\author[1]{Ruipeng Dong\thanks{Dong and Li are co-first authors.}}
%\author[2]{Daoji Li\samethanks}
%\author[1]{Zemin Zheng\thanks{Corresponding author. Email: zhengzm@ustc.edu.cn.}}
%
%\affil[1]{\footnotesize{International Institute of Finance, The School of Management, \protect\\ University of Science and Technology of China}}
%\affil[2]{\footnotesize{Department of Information Systems and Decision Sciences, \protect\\ California State University, Fullerton}}
%\date{April 11, 2021}

\title{Parallel integrative learning for large-scale multi-response regression with incomplete outcomes
	\thanks{Ruipeng Dong is Ph.D. candidate, International Institute of Finance, The School of Management, University of Science and Technology of China, Hefei, Anhui, 230026, China (E-mail:drp@mail.ustc.edu.cn). Daoji Li is Assistant Professor, Department of Information Systems and Decision Sciences, California State University, Fullerton, CA, 92831 (E-mail:dali@fullerton.edu). Zemin Zheng is Professor, International Institute of Finance, The School of Management, University of Science and Technology of China, Hefei, Anhui, 230026, China (E-mail:zhengzm@ustc.edu.cn). Li was supported by 2020 individual Award (0358220) from the Innovative Research and Creative Activities Grant at California State University, Fullerton. Zheng was
	supported by National Natural Science Foundation of China (Grants 72071187, 11671374, 71731010, and 71921001) and Fundamental Research Funds for the Central Universities (Grants WK3470000017 and WK2040000027). }}
\author[1]{Ruipeng Dong}
\author[2]{Daoji Li}
\author[1]{Zemin Zheng}

\affil[1]{University of Science and Technology of China}
\affil[2]{California State University, Fullerton}
\date{April 11, 2021}
\maketitle

\begin{abstract}
	Multi-task learning is increasingly used to investigate the association structure between multiple responses and a single set of predictor variables in many applications. In the era of big data, the coexistence of incomplete outcomes, large number of responses, and high dimensionality in predictors poses unprecedented challenges in estimation, prediction and computation. In this paper, we propose a scalable and computationally efficient procedure, called PEER, for large-scale multi-response regression with incomplete outcomes, where both the numbers of responses and predictors can be high-dimensional. Motivated by sparse factor regression, we convert the multi-response regression into a set of univariate-response regressions, which can be efficiently implemented in parallel. Under some mild regularity conditions, we show that PEER enjoys nice sampling properties including consistency in estimation, prediction, and variable selection. Extensive simulation studies show that our proposal compares favorably with several existing methods in estimation accuracy, variable selection, and computation efficiency.
\end{abstract}


\section{Introduction}

Multi-task learning has been widely used in %various disciplines, 
various fields, such as bioinformatics %\citep{wang2007, kim2009multivariate, hilafu2020sparse},
\citep{kim2009multivariate, hilafu2020sparse},
%computational biology \citep{kim2009multivariate}, 
econometrics \citep{fan2019generalized}, social network analysis \citep{zhu2020multivariate}, and  recommender systems \citep{shenxiaotong2016}, when one is interested in
uncovering the association between multiple responses and a single set of predictor variables.
Multi-response regression is one of the most important tools in multi-task learning. For example, investigating the relationship between several measures of health of a patient (i.e., cholesterol, blood pressure, and weight) and eating habits of this patient, or simultaneously predicting asset returns for several companies via %the vector auto-regressive model, 
vector autoregression models, both result in multi-response regression problems.

In the high-dimensional setting where the number of
predictors is large, it is challenging to infer the association between predictors and responses
because the responses may depend on only a subset of predictors. To address this issue and recover sparse response-predictor associations, many regularization methods for multi-response regression models have been proposed; see, for example, \cite{rothman2010sparse},
%\cite{bunea2011optimal}, 
\cite{bunea2011optimal, bunea2012joint}, \cite{lisha2012}, \cite{chen2012},  \cite{chen2016note}, \cite{uematsu2019}, %\cite{zou2020estimation}
and the references therein.
In particular, \cite{chen2012} and \cite{chen2016note} have proposed sparse reduced-rank regression approaches, which combine the regularization and reduced-rank regression techniques \citep{izenman1975reduced, velu2013multivariate}, and \cite{uematsu2019}
suggested the method of sparse orthogonal factor regression via the sparse singular value decomposition with
orthogonality constrained optimization to find the underlying association networks.

In the era of big data, the coexistence of missing values, large number of responses, and high dimensionality in predictors is increasingly common in many
%modern big data 
applications. When both numbers of responses and predictors are large,  the aforementioned methods may become inefficient because
they are computationally intensive. %, mainly due to the simultaneous presence of orthogonality constraints and sparsity regularization. 
In addition, these methods are not applicable to incomplete data
because they mainly focus on full data problems. To obtain scalable estimation of sparse reduced-rank regression,
some approaches based on sequential estimation techniques have been developed in recent years. To name a few, \citet{mishra2017}
proposed a sequential extraction
procedure for model estimation, which extracts unit-rank factorization one by one in a sequential fashion,
each time with the previously extracted components removed from the current response
matrix. Although \cite{mishra2017} also considered extensions to incomplete outcomes, they did not provide the theoretical justification for the case with incomplete outcomes.
In addition, the sequential steps in %the suggested 
their procedure may result in the error accumulation.
Alternatively, \citet{zheng2019} converted the sparse and low-rank regression problem to a sparse generalized eigenvalue problem and recovered the underlying coefficient matrix %sequentially
in a similar sequential fashion. Although this method has been shown to enjoy desirable theoretical properties, it cannot be applied directly to missing data.

In this paper, we propose a new methodology of parallel integrative learning regression (PEER) for large-scale multi-task learning with incomplete outcomes,
where both responses and predictors are possibly of high dimensions. PEER is a novel two-step procedure, where in the first step we consider a constrained optimization and use an iterative singular value thresholding algorithm to obtain some initial estimates, and then in the second step we convert the multi-response regression into a set of univariate-response regressions, which can be efficiently implemented in parallel.

The major contributions of this paper are threefold. First,
the proposed procedure PEER provides a scalable and computationally efficient approach to large-scale multi-response regression models with incomplete outcomes. PEER can uncover the association between multiple responses and a single set of predictor variables while simultaneously achieving
dimension reduction and variable selection. Second, our procedure PEER %solves 
addresses the error accumulation problem in existing sequential estimation approaches by converting the %sparse 
multi-response regression into a set of parallel univariate-response regressions. Third, we provide theoretical guarantees for %the suggested method 
PEER by establishing oracle inequalities in estimation and prediction. Our theoretical analysis shows that PEER can consistently estimate the singular vectors, latent
factors as well as the regression coefficient matrix, and accurately predict the multivariate response vector under mild conditions. To the best of our knowledge, there is no existing theoretical result on large-scale multi-response regression with incomplete outcomes. Our theoretical results are new to the literature.

The rest of this paper is organized as follows. Section \ref{sec:model-algorithm} introduces the  model setting and our new procedure PEER.
Section \ref{sec:theory} establishes non-asymptotic properties of PEER in high dimensions.
Section \ref{sec: sim} illustrates the advantages of our method via extensive simulation
studies. Section \ref{sec:yeast} presents the results of a real data example. Section \ref{sec:discussion} concludes with some discussions. All the proofs are relegated to the Appendix.

\section{Model and Methodology}\label{sec:model-algorithm}

In this section, we first introduce our model setting and
briefly review sparse orthogonal factor regression framework
for high-dimensional multi-response regression models. We then present our new approach PEER.

\subsection{Model setting and sparse orthogonal factor regression}\label{data-model}

Given $n$ observations of the vector of responses $\y\in\R^{q}$ and vector of predictors $\x\in\R^{p}$, we consider the following multi-response regression model
\begin{align}\label{model-matrix}
	\Y = \X\C^* + \E,
\end{align}
where $\Y = \left(\y_1,\dots,\y_n\right)\trans\in\R^{n\times q}$ is the response matrix, $\X = \left(\x_1,\dots,\x_n\right)\trans \in \R^{n\times p}$ is the design matrix,
$\C^*\in\R^{p\times q}$ is the  regression coefficient matrix, %with $\rank{\C^*}=r^*\leq \min\{p, q\}$,
and $\E = \left(\e_1,\dots,\e_n\right)\trans\in\R^{n\times q}$ is the error matrix. We consider fixed design in this paper and assume that responses and predictors are centered so that there is no intercept term. Without loss of generality, we assume that each  column of $\X$ is rescaled to have an $\ell_2$-norm $n^{1/2}$.

Similar to \cite{mishra2017}, \cite{uematsu2019} and %\cite{uematsu2019} 
\cite{zheng2019}, we consider  model %\eqref{eq: model-0} 
\eqref{model-matrix}
from a latent factor regression point of view. More specifically,
%assume $\rank{\C^*}=r^*$ with the matrix rank $r^*\leq \min\{p, q\}$. 
assume the matrix rank of $\C^*$ is $r^*$ with $r^*\leq \min\{p, q\}$.
We can write
$\C^*=\U^*\D^*\V\strans$ where $\U^*=\left(\u_1^*,\dots,\u_{r^*}^*\right)\in\R^{p\times r^*}$, $\V^*=\left(\v_1^*,\dots,\v_{r^*}^*\right)\in\R^{q\times r^*}$,
and $\D^*=\text{diag}\left\{d_1^*,\dots,d_{r^*}^*\right\}$ is an $r^*\times r^*$ diagonal matrix with singular values
$d_1^*\geq d_2^*\geq \cdots\geq d_{r^*}^*>0$.
To avoid redundancy, it is desirable to make these latent factors uncorrelated by imposing the constraint %that
$\mbox{cov}(\U\strans\x)=\I_{r^*}$. This leads to $\U\strans\bGamma\U^*=\I_{r^*}$, where $\bGamma=\mbox{cov}(\x)$ is the covariance matrix of %the predictor vector 
the vector of predictors
$\x$. Similar to the factor analysis, to ensure the parameter identifiability, we also require $\V\strans\V^*=\I_{r^*}$.  Thus $\C^*$ admits the following representation
\begin{align*}
	\C^*=\U^*\D^*\V\strans\quad\mbox{subject to} \quad\U\strans\bGamma\U^*=\V\strans\V^* =\I_{r^*}.
\end{align*}
Note that the population covariance matrix $\bGamma$ is unknown. Using $n^{-1}\X^T\X$, the Gram matrix of the predictors, to replace its population counterpart $\bGamma$, we have the following decomposition
\begin{align}\label{eq: C-SVD-decomp}
	\C^*=\U^*\D^*\V\strans\,\,\mbox{subject to} \,\,\left(\frac{1}{\sqrt{n}}\X\U^*\right)\trans\left(\frac{1}{\sqrt{n}}\X\U^*\right)=\V\strans\V^* =\I_{r^*}.
\end{align}
Thus we can write the coefficient matrix $\C^*$ as
\begin{align*}%\label{eq: C}
	\C^*=\sum_{k=1}^{r^*}d_k^*\u_k^*\v_k\strans
	=\sum_{k=1}^{r^*}\C^*_k,
\end{align*}
where $\C^*_k=d_k^*\u_k^*\v_k\strans$ is
%the layer $k$ unit rank matrix of $\C^*$, 
the unit rank matrix corresponding to
the $k$th layer of $\C^*$,
$\u^*_k$ and $\v^*_k$
are the $k$th column of $\U^*$ and $\V^*$, respectively, and $d^*_k$ is the $k$th diagonal element of $\D^*$.

The decomposition \eqref{eq: C-SVD-decomp} gives a latent factor regression model %$\Y=\X\U^*\D^*\V\strans + \E$ 
\begin{equation}\label{model:decomposition}
	\begin{aligned}
		 & \Y=\X\U^*\D^*\V\strans + \E                                                                                                    \\
		 & \mbox{subject to} \,\,\left(\frac{1}{\sqrt{n}}\X\U^*\right)\trans\left(\frac{1}{\sqrt{n}}\X\U^*\right)=\V\strans\V^* =\I_{r^*}
	\end{aligned}
\end{equation}
with $r^*$ latent factors, where $\X\u_{k}^*$ is the $k$th latent factor, $\u_k^*$ gives the weights for
constructing the $k$th latent factor,
$\v_k^*$ describes the impacts of the $k$th latent factor on the response variables, and
$d_k^*$ indicates the importance of the $k$th factor for $k=1, 2, \dots, r^*$.
Each left singular vector $\u_{k}^*\in \R^{p}$ is assumed to be sparse. Without loss of generality, here we assume
that $\rank{\X\C^*}=\rank{\C^*}$ since the redundant part of $\C^*$ can be removed if $\rank{\X\C^*}<\rank{\C^*}$ such that it reflects the true number of latent factors.
Thanks to the orthogonality of $\X\U^*$, the sample latent factors are uncorrelated with each other. The low-rank structure imposed on the unknown coefficient matrix
$\C^*$ yields that  all responses can be predicted by a relatively small set of common factors. On the other hand, under the sparsity assumption of $\u_k^*$, each latent factor depends  on only a subset of original predictors which facilitates the interpretation of model with the high-dimensional data.

However, unlike \cite{mishra2017} and \cite{uematsu2019}, we do not require the right singular vectors $\v_k^*$'s to be sparse.
In this paper, we consider large-scale multi-response regression models with incomplete outcomes where the response matrix $\Y$ may not be fully observed and
both the numbers of responses and predictors can be high-dimensional.
Denote by $\M$ the index set
of all observed values in the response matrix $\Y$, that is,
\begin{align}\label{eq: SetM-def}
	\M=\{(i, j): y_{ij}\,\,\text{is observed},\,\, 1\leq i\leq n, 1\leq j\leq q\}.
\end{align}
We will focus on coefficient matrix estimation and variable selection.

Next, we introduce some notation and definitions which will be used  throughout the paper.
Denote by $a\wedge b=\min\{a, b\}$ and $a\vee b=\max\{a, b\}$.
For any vector $\a=(a_i)$, denote by $\|\a\|_0$ the number of non-zero entries in $\a$, and let $\|\a\|_1$, $\|\a\|_2$, and $\|\a\|_{\infty}$ be the $\ell_{1}$-norm, $\ell_{2}$-norm,
and $\ell_{\infty}$-norm, respectively, which are defined as $\|\a\|_1=\sum_i|a_i|$, $\|\a\|_2=\left(\sum_ia_i^2\right)^{1/2}$, and
$\|\a\|_{\infty}=\max_i|a_i|$. For any matrix $\A=(a_{ij})$, denote by
$\norm{\A}_{F}=\left(\sum_{i, j}a^2_{ij}\right)^{1/2}$, $\norm{\A}_{op}=\max_{\u\neq 0}\norm{\A\u}_2/\norm{\u}_2$, and $\norm{\A}_{\max}=\max_{i,j}\abs{a_{ij}}$ the  Frobenius  norm,  the  operator norm,  and  the  entrywise  maximum  norm, respectively.
In addition, we use $\rank{\A}$ to denote the rank of $\A$, and $d_k(\A)$ to denote the $k$th largest singular value of $\A$.
Let $\proj_\M(\A)$ denote the projection of $\A$ onto $\M$,
which is
the matrix with the observed elements of $\A$ preserved, and the missing entries replaced with $0$. Then $\A =\proj_\M(\A) + \proj_{\M^c}(\A)$. For an index set $J\subset\{1, \cdots, p\}$, denote by $J^c$ the complement of a set $J$ and $\bdelta_{J}$ the subvector of $\bdelta\in\R^{p}$ formed by components in $J$. Let $\abs{J}$ be the cardinality of $J$. Finally, $a\lesssim b$ means that $a$ is less than $cb$
with some positive constant $c$.

\subsection{Parallel integrative learning via PEER}\label{sec: method-PEER}

In this subsection, we will introduce our new method PEER.
Recall that our goal is to accurately
estimate not only the low-rank coefficient matrix $\C^*$
but also
$\u^*_k$, $\v^*_k$ and $d^*_k$ such that we can recover the latent factors, the significant predictors, and their impacts.
Motivated by the decomposition in \eqref{model:decomposition}, %when the response matrix $\Y$ is fully observed, 
we introduce a two-step procedure, where in the first step we consider a constrained optimization and propose an iterative singular value thresholding algorithm to %estimate $\D^*$ and $\V^*$, 
obtain some initial estimates,
and then in the second step we employ
a scalable and efficient approach to
estimate $\u^*_k$, $\v^*_k$, $d^*_k$,
%$\U^*$, $\V^*$, $\D^*$, 
and $\C^*$ and select important predictors for each latent factor.

The first step of our method PEER is to consider
the following constrained optimization problem
\begin{equation}\label{opt:low-rank-approx-incom}
	\begin{aligned}
		(\widetilde{\D},\widetilde{\Z},\widetilde{\V})
		 & = \underset{\D,\Z,\V}{\arg\min}~m^{-1}\norm{\proj_\M(\Y) - \proj_\M(\Z\D\V\trans)}_F^2, \\
		 & ~\text{subject to}~ \Z\trans\Z = \V\trans\V = \I_r,
	\end{aligned}
\end{equation}
where $m=|\M|$ is the cardinality of the index set $\M$ in \eqref{eq: SetM-def},  $\widetilde{\D}=\text{diag}\left\{\widetilde{d}_{1},\dots,\widetilde{d}_{r}\right\}$ is an $r\times r$ diagonal matrix with $1\leq r\leq \min\{p, q\}$,
$\widetilde{\Z}=\left(\widetilde{\z}_1,\dots,\widetilde{\z}_{r}\right)\in\R^{n\times r}$ and
$\widetilde{\V}=\left(\widetilde{\v}_{1},\dots,\widetilde{\v}_{r}\right)\in\R^{q\times r}$. Without loss of generality, we assume that the singular values in $\widetilde{\D}$ are placed in descending order.  As pointed out by \cite{uematsu2019}, %although it is always possible to set $r=\min\{p, q\}$ 
when prior knowledge of the
rank $r^*$ is not available, it is often sufficient in practice to take an $r$ such that it is slightly larger than the expected rank (estimated by some similar procedure such as in \cite{bunea2011optimal}).
The solution $(\widetilde{\D},\widetilde{\Z},\widetilde{\V})$
from \eqref{opt:low-rank-approx-incom} will be used as our initial estimates to estimate
$\u^*_k$, $\v^*_k$, $d^*_k$, and $\C^*$
in the second step of PEER.

We use an iterative singular value thresholding algorithm to solve the optimization problem \eqref{opt:low-rank-approx-incom}. To ease the presentation, for any matrix $\A\in\R^{n\times p}$, let $\A=\L\S{\bf R}\trans$ be the singular value decomposition of $\A$ with a diagonal matrix $\S$ including all singular values of $\A$. We define $\T(\A; r) = \L\widetilde{\T}(\S; r){\bf R}\trans$, where
$\widetilde{\T}(\S; r)$ is the diagonal matrix
with the first $r$ largest entries on the main diagonal of $\S$ preserved, and other entries on the main diagonal of $\S$ replaced with $0$. In other words, the $r$ largest singular values of $\T(\A; r)$ are the same as those of
$\A$. The details of the iterative singular value thresholding algorithm for solving the optimization problem \eqref{opt:low-rank-approx-incom} are provided in Algorithm
\ref{algorithm:low-rank-approximate}.

\begin{algorithm}[!htb]
	\caption{Iterative singular value thresholding algorithm}
	\label{algorithm:low-rank-approximate}
	\begin{algorithmic}[1]
		\Require response matrix $\Y\in\R^{n\times q}$, rank $r$ and tolerance parameter $\epsilon$.
		\State Update $\Y$ by replacing the missing values by the column averages
		of observed entries of $\Y$ and then set $\A^{\text{new}}=\Y$.
		\Repeat
		\State $\A^{\text{old}}\leftarrow \A^{\text{new}}$
		\State $\A^{\text{new}} \leftarrow \T(\Y;r)$
		\State $\Y \leftarrow \proj_{\M}(\Y) + \proj_{\M^c}(\A^{\text{new}})$
		\Until{$\norm{\A^{\text{new}} - \A^{\text{old}}}_F/\norm{\A^{\text{old}}}_F \leq \epsilon$}
		\State Compute the SVD of $\A^{\text{new}}$ such that $\A^{\text{new}} = \widetilde{\Z}\widetilde{\D}\widetilde{\V}\trans$ with
		$\widetilde{\Z}\in\R^{n\times r}$,
		$\widetilde{\V}\in\R^{q\times r}$
		and $\widetilde{\D}=\text{diag}\{\widetilde{d}_{1},\dots,\widetilde{d}_{r}\}$.
		\State \Return $\widetilde{\Z}$, $\widetilde{\D}$ and $\widetilde{\V}$.
	\end{algorithmic}
\end{algorithm}

The second step of PEER is to estimate $\u^*_k$, $\v^*_k$, $d^*_k$, and $\C^*$. Once we obtain $\widetilde{\D}$ and $\widetilde{\V}$ from \eqref{opt:low-rank-approx-incom}, we can estimate $\D^*$ and $\V^*$ by $\widehat{\D}=n^{-1/2}\widetilde{\D}$ and $\widehat{\V}=\widetilde{\V}$, respectively.
In other words, $d_k^*$ and $\v_k^*$ are estimated by  $\widehat{d}_k=n^{-1/2}\widetilde{d}_k$ and  $\widehat{\v}_k=\widetilde{\v}_k$, respectively, for $k=1, \cdots, r$.
Here we rescale $\widetilde{\D}$  because the singular values of $n^{-1/2}\Y$ are
$n^{-1/2}$ times of the singular values of $\Y$.
Note that $\widetilde{\z}_k$ is an estimate of the matrix $n^{-1/2}\X\u^*_k$.
We can estimate $\u_k^*$ by solving the univariate
response Lasso regression
\begin{align}\label{eq: uk-optim}
	\widehat{\u}_k = \underset{\u_k\in\R^p}{\arg\min}~n^{-1}\norm{\sqrt{n}\,\widetilde{\z}_k - \X\u_k}_2^2 + \lambda_k\norm{\u_k}_1,
\end{align}
where $\lambda_k$ is a regularization parameter and can be tuned by cross-validation or certain information criterion.
The univariate response
Lasso regression \citep{tibshirani1996} has been studied extensively in the literature
and many efficient algorithms have been proposed
for solving it. See, for examples, \cite{efron2004least}, \cite{zhao2006model}, \cite{friedman2007pathwise},  \cite{bunea2007sparsity}, \cite{van2008high}, \cite{wu2008coordinate}, \cite{bickel2009},   %\cite{friedman2010regularization},
%\cite{fan2013asymptotic}, 
and the references therein.
Once each $(\widehat{d}_k,\widehat{\u}_k,\widehat{\v}_k)$ is obtained for
$k=1, \cdots, r$ with a given $r$, one can use cross validation or other criterion to estimate the true  rank $r^*$. In this paper, we propose a thresholding procedure to estimate $r^*$ in Theorem \ref{th3}.
Denote by $\widehat{r}$ the estimated rank. Then we can estimate regression coefficient matrix $\C^*$ by $\widehat{\C}=\sum_{k=1}^{\widehat{r}}\widehat{\C}_k$ with $\widehat{\C}_k=\widehat{d}_k\widehat{\u}_k\widehat{\v}_k^T$.
This leads to our complete algorithm for PEER, which is described
in Algorithm \ref{algorithm2: PEER}.

\begin{algorithm}[!htb]
	\caption{PEER}
	\label{algorithm2: PEER}
	\begin{algorithmic}[1]
		\Require response matrix $\Y\in\R^{n\times q}$, design matrix $\X\in\R^{n\times p}$, initial rank $r$, and tolerance parameter $\epsilon$
		\State obtain $\widetilde{\Z}$,
		$\widetilde{\D}$ and $\widetilde{\V}$ from \eqref{opt:low-rank-approx-incom} using Algorithm \ref{algorithm:low-rank-approximate}. \Comment{Step one}
		\ForAll {$k=1, \cdots, r$}\Comment{Step two}
		\State {obtain $\widehat{\u}_k$ from \eqref{eq: uk-optim} using $\widetilde{\z}_k$ and $\X$ where $\widetilde{\z}_k$ is the $k$th column of $\widetilde{\Z}$}
		\State {$\widehat{\v}_k\leftarrow\widetilde{\v}_k$ where $\widetilde{\v}_k$ is the $k$th column of $\widetilde{\V}$}
		\State {$\widehat{d}_k\leftarrow n^{-1/2}\widetilde{d}_k$ where $\widetilde{d}_k$ is the $k$th main diagonal element of $\widetilde{\D}$}
		\EndFor
		\State use \eqref{eq: rank-hat} to estimate the true rank $r^*$ and obtain $\widehat{r}$
		\State \Return	$\widehat{\C}=\sum_{k=1}^{\widehat{r}}\widehat{\C}_k$ with $\widehat{\C}_k=\widehat{d}_k\widehat{\u}_k\widehat{\v}_k^T$
	\end{algorithmic}
\end{algorithm}

We remark that although the $L_1$ penalty is used in \eqref{eq: uk-optim}, one can use any favorite variable selection method in the second step of PEER, for example, Adaptive Lasso \citep{zou2006adaptive}, SCAD \citep{fan2001variable},
SICA \citep{lv2009unified}, and MCP
\citep{zhang2010nearly}, among many others. See also \cite{fan2013asymptotic} for the asymptotic equivalence
of various regularization methods.

Recall that there are $p$ predictors and $q$ responses in our model \eqref{model:decomposition}.
Thus, the original problem of model fitting and variable selection in our model %\eqref{model:decomposition}
involves a large-scale optimization problem when both $p$ and $q$ are large. However, the algorithm used in the first step of our procedure only depends on basic matrix operations that can be efficiently %calculated
implemented in high performance computing
devices. Thanks to the first step, the problem of estimating $\U^*$ %, $\V^*$, and $\D^*$ 
and selecting important predictors for each latent factor in the second step can be recast as $r$ univariate response Lasso regressions, which can be efficiently implemented in parallel. In addition, after obtaining initial estimates $\widetilde{\v}_k$ and $\widetilde{d}_k$, estimating $\v^*_k$ and $d^*_k$ in the second step is straightforward. % and can be efficiently parallelized in the second step. 
See Figure \ref{fig:inllustration} for an illustration.

\tikzstyle{block} = [draw,minimum size=2em]
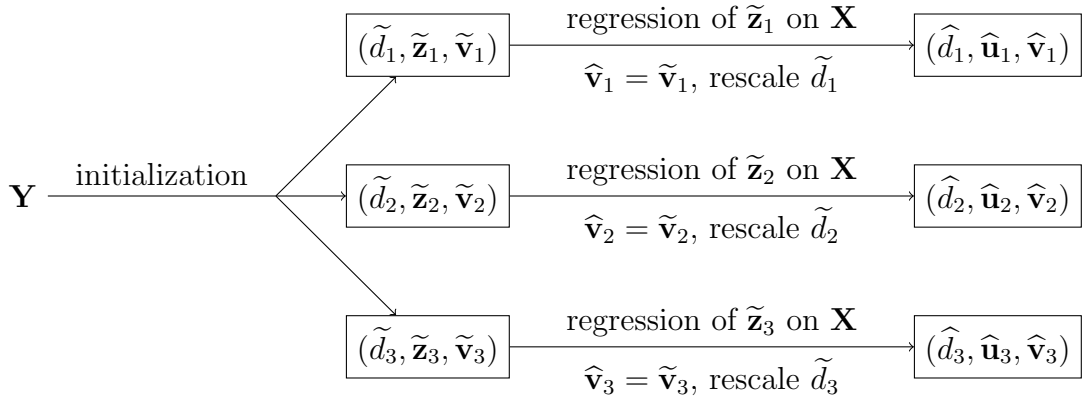
\begin{figure}[!htb]
	\centering
	\begin{tikzpicture}
		\draw (-3,0) node[left]{$\Y$} -- node[pos=0.5,above,sloped]{initialization}(0,0);
		\foreach \y in {1,2,3} {
				\node[block] at (2,4-\y*2) (input\y) {$(\widetilde{d}_\y,\widetilde{\z}_\y,\widetilde{\v}_\y)$};
				\node[block] at (9.5,4-\y*2) (output\y) {$(\widehat{d}_\y,\widehat{\u}_\y,\widehat{\v}_\y)$};
				\draw[->] (0,0) -- (input\y);
				\draw[->] (input\y) -- node[pos=0.5,above,sloped]{ regression of $\widetilde{\z}_\y$ on $\X$} node[pos=0.5,below,sloped]{$\widehat{\v}_\y=\widetilde{\v}_\y$, rescale $\widetilde{d}_\y$} (output\y);
			}
	\end{tikzpicture}
	\caption{An illustration of the parallel computation with $r=3$.}\label{fig:inllustration}
\end{figure}

%We would like to remark that 
Our method PEER can be simplified to handle the case
with full data
where the response matrix $\Y$ is fully observed. In this case, there is no need to use Algorithm
\ref{algorithm:low-rank-approximate} in the first step to obtain initial estimates
$\widetilde{\Z}$, $\widetilde{\D}$ and $\widetilde{\V}$. In fact, when the response matrix $\Y$ is fully observed, one can directly consider the singular value decomposition $\Y = \widetilde{\Z}\widetilde{\D}\widetilde{\V}\trans$
and use $\widetilde{\Z}$, $\widetilde{\D}$ and $\widetilde{\V}$ as initial estimates for the second step of PEER to estimate $\u^*_k$, $\v^*_k$, $d^*_k$, and $\C^*$.

\section{Theoretical properties}\label{sec:theory}

In this section, we investigate the theoretical properties of PEER. We first list some mild regularity conditions that facilitate our technical analysis.

\subsection{Technical conditions}

\begin{condition}\label{cond1}
	The error matrix $\E$ has independent sub-Gaussian entries $e_{ij}$ with $\mathbb{E}(e_{ij})=0$ and $\var(e_{ij})=\sigma^2>0$.
\end{condition}

\begin{condition}\label{cond2}
	There exists some constant $\gamma_{d} > 0$ such that non-zero singular values satisfy $q^{-1/2}(d_{k}^* - d_{k+1}^*) \geq \gamma_{d}$ for all $1 \leq k \leq r^{*}-1$. In addition,  $d_{k}^*=O(\sqrt{q})$ for $1 \leq k \leq r^{*}$.
\end{condition}

\begin{condition}\label{cond4}
	There exists certain sparsity level $s$ with a positive constant $\rho_{l}$ such that
	\begin{align*}
		\underset{\bdelta\in\mathbb{R}^p}{\inf} \left\{\frac{\norm{\X\bdelta}_{2}^{2}}{n(\norm{\bdelta_{J}}_{2}^{2}\vee\norm{\bdelta^{(1)}_{J^c}}_{2}^{2})}:
		\abs{J} \le s, ~
		\norm{\bdelta_{J^c}}_{1} \le 3 \norm{\bdelta_{J}}_{1}\right\} \geq \rho_{l},
	\end{align*}
	where $\bdelta^{(1)}_{J^c}$ is a
	subvector of $\bdelta_{J^c}$ consisting of the $s$ largest components in magnitude.
\end{condition}

\begin{condition}\label{cond3}
	Let $\pi_{ij}$ be the probability of $y_{ij}$ being observed for
	all $1\leq i \leq n$ and $1\leq j \leq q$. Then there exist some constants $\mu\geq 1$ and $\nu \geq 1$ such that
	\begin{equation*}
		\begin{aligned}
			 & \min_{1\leq i \leq n, 1\leq j \leq q}~\pi_{ij} \geq 1/(\mu nq)\,\,\,\mbox{and}\,\,\,
			\max_{1\leq i \leq n, 1\leq j \leq q}\left\{\sum_{i=1}^n\pi_{ij},\sum_{j=1}^q\pi_{ij}\right\} \leq \nu/(n \wedge q).
		\end{aligned}
	\end{equation*}
\end{condition}

\begin{condition}\label{cond5}
	There exists a positive constant $L$ such that $\norm{\X\C^*}_{\max} \leq L$.
	We also assume
	$m\geq \max\{\nu^{-1}(n\wedge q)\log^{3}(n+q),\, r(n\vee q)\log(n+q)\}$.
\end{condition}

Condition \ref{cond1} is a common assumption to control the tail behavior of the random errors. Gaussian distribution
and distributions with bounded support are two special examples of sub-Gaussian distribution.

Condition \ref{cond2} requires strict
separation among the singular values which can ensure that the first $r^*$ left singular vectors are distinguishable.
Condition \ref{cond2} also assumes a spiked eigen-structure which is $d_{k}^*=O(\sqrt{q})$. This rate is reasonable since we do not impose sparsity on the columns of $\C^*$. Similar assumptions can be found in the literature; see, for example, \citet{zheng2019, zheng2021}.

Condition \ref{cond4} combines
the restricted eigenvalue (RE) assumptions in \cite{bickel2009}, which has been been commonly used to establish the oracle inequalities for the Lasso and Dantzig selector \citep{candes2007dantzig}.

Condition \ref{cond3} puts constraints on %the sampling probabilities.
the probabilities of entries of the response matrix %$\Y$ 
being observed.
To be specific, the first inequality requires the sampling probability of each entry of $\Y$ is bounded below by a positive constant while the second one
ensures that neither a row nor a column should be sampled far more frequently than the others. When $\mu=\nu=1$, the condition corresponds to the special case of uniform sampling. The same condition has also been used in
\cite{klopp2014}, \cite{lafond2015low}, and \cite{luo2018}.

Condition \ref{cond5} is assumed mainly for theoretical analysis. The first part of Condition \ref{cond5} is not restrictive because we consider fixed design $\X$ in this paper. The second part of Condition \ref{cond5} imposes a lower bound on the number of observed entries of $\Y$. Intuitively, the estimation may fail when the number of observed entries is too small.

\subsection{Main results}

For the theoretical analysis purpose, we reformulate the optimization \eqref{opt:low-rank-approx-incom} as
\begin{align}\label{opt:matrix-completion}
	\widetilde{\Y} = \underset{\A\in\mathcal{Y}}{\arg\min}~m^{-1}\norm{\proj_\M(\Y) - \proj_\M(\A)}_F^2,
\end{align}
where $\mathcal{Y}=\left\{\A\in\R^{n\times q}:\norm{\A}_{\max}\leq L,~ r^* \leq \rank{\A} = r\right\}$. With $\widetilde{\Y}$, we consider its singular value decomposition
$\widetilde{\Y}=\widetilde{\Z}\widetilde{\D}\widetilde{\V}\trans$, where $\widetilde{\Z}=(\widetilde{\z}_{1},\dots,\widetilde{\z}_{r})\in\R^{n\times r}$,
$\widetilde{\V}=\left(\widetilde{\v}_{1},\dots,\widetilde{\v}_{r}\right)$ and $\widetilde{\D}=\text{diag}\left\{\widetilde{d}_{1},\dots,\widetilde{d}_{r}\right\}$.
Compared with \eqref{opt:low-rank-approx-incom}, the only difference is the constraint $\norm{\A}_{\max}\leq L$ in \eqref{opt:matrix-completion}. This constraint is mainly for theoretical analysis since it is not used in our practical implementation for PEER; see
Algorithm \ref{algorithm2: PEER} in Section \ref{sec: method-PEER} for details.
We first introduce the following lemma, which establishes the consistency of top-$r^{*}$ latent factors, top-$r^{*}$ right singular vectors, and top-$r^{*}$ singular values in the first step of PEER.
\begin{lemma}\label{incomp:lem2}
	Under Conditions \ref{cond1}, \ref{cond2}, \ref{cond3}, and \ref{cond5}, we have that
	\begin{align*}
		\norm{\widetilde{\z}_k - n^{-1/2}\X\u_k^*}_2 \lesssim \gamma_{d}^{-1} B_n, \,\,\,\,
		\norm{\widetilde{\v}_k -\v_k^*}_2 \lesssim \gamma_{d}^{-1} B_n,\,\,\,\,
		\frac{\abs{\widetilde{d}_k - \sqrt{n}d_k^*}}{\sqrt{nq}} \lesssim B_n,
	\end{align*}
	hold uniformly over $k=1, \cdots, r^*$ with probability at least $1-(n+q)^{-1}$,
	where $B_n$ is given by
	\begin{align}\label{eq: Bn-def}
		B_{n} = \max\left[L\mu^{1/2}\left\{\frac{\log(n+q)}{m}\right\}^{1/4},~
		\mu(\sigma\vee L)\nu^{1/2}\sqrt{\frac{r(n \vee q)\log(n+q)}{m}}\,\right].
	\end{align}
\end{lemma}

The results of Lemma \ref{incomp:lem2} are the bases of our two-step procedure for estimating $\u^*_k$, $\v^*_k$, $d^*_k$, and $\C^*$. The following theorem establishes the estimation and prediction bounds of PEER with incomplete outcomes. It also demonstrates that our method PEER enjoys oracle inequalities with
non-asymptotic convergence rates for top-$r^*$ layers. % when the response matrix $\Y$ cannot be fully observed.

\begin{theorem}	[Estimation and prediction bounds with incomplete outcomes]\label{im:th}
	Suppose that Condition \ref{cond1}--\ref{cond5} holds and the sparsity level $s\geq \max\limits_{1\leq k\leq r^*}s_k$ with $s_k=\norm{\u_k^*}_0$. Choose $\lambda_k = 4c\gamma_{d}^{-1}B_{n}$ with $c$ some positive constant, where $B_n$ is defined in \eqref{eq: Bn-def}. Then with probability at least $1-(n+q)^{-1}$, the following inequalities
	\begin{align*}
		\norm{\widehat{\u}_k - \u_k^*}_2 \lesssim (\gamma_{d}\rho_l)^{-1}
		\sqrt{s_k}B_n, \quad\quad %\label{im: uk-bound}
		 & \norm{\widehat{\v}_k - \v_k^*}_2 \lesssim \gamma_{d}^{-1} B_n,                                                    %\label{im: vk-bound}
		\\
		q^{-1/2}\abs{\widehat{d}_k - d_k^*} \lesssim B_n, \quad\quad %\label{im: dk-bound}
		 & q^{-1/2}\norm{\widehat{\C}_{k}-\C^*_{k}}_{F} \lesssim (\gamma_{d}\rho_l)^{-1}
		\sqrt{s_k}B_n,  %\label{im: Ck-bound}
		\\
		\frac{\norm{\X(\widehat{\u}_k - \u_k^*)}_2}{\sqrt{n}} \lesssim  \frac{\sqrt{s_k}B_n}{\gamma_{d}\sqrt{\rho_l}},\quad\quad %\label{im: Xuk-bound}
		 & \frac{\norm{\X(\widehat{\C}_{k}-\C^*_{k})}_{F}}{\sqrt{nq}} \lesssim \frac{\sqrt{s_k}B_n}{\gamma_{d}\sqrt{\rho_l}} %\label{im: XCk-bound}
	\end{align*}
	hold uniformly over $k=1, \cdots, r^*$.
\end{theorem}

Theorem \ref{im:th} presents the oracle inequalities
and establishes non-asymptotic convergence rates for top-$r^*$ layers when the response matrix $\Y$ may not be fully observed. To be specific, it gives the uniform estimation error bounds for top-$r^*$ left singular vectors $\u_k^*$, right singular vectors $\v_k^*$, singular values $d_k^*$, unit rank matrices $\C^*_{k}$, latent factors $\X\u_k^*$, and the uniform prediction error bounds of the top-$r^*$ layers $\X\C_k^*$. The factor $\sqrt{s_k}$ in the estimation error bounds for $\u_k^*$ and $\X\u_k^*$ reflects the sparsity constraint as there are $s_k$
non-zero components in $\u_k^*$ for each $k=1, \cdots, r^*$.
Note that the bounds on right singular vectors $\v_k^*$ do not involve the factor $\sqrt{s_k}$ since there is no sparsity constraint on $\v_k^*$.

Our results in Theorem \ref{im:th} are new to the literature. As mentioned in the Introduction, there is no existing theoretical result when the response matrix $\Y$ cannot be fully observed.
Although \cite{mishra2017} considered extensions to incomplete data,
they did not provide corresponding theoretical justification. To the best of our knowledge, Theorem \ref{im:th} provides the first formal theoretical result on large-scale multi-response regression with incomplete outcomes. In addition,
our error bounds for incomplete outcomes %in Theorem \ref{im:th} 
are all non-asymptotic while most existing results for complete data are asymptotic. For example, \cite{mishra2017} and \cite{zheng2019} have focused on complete data where the response matrix $\Y$ is fully observed and all corresponding results are asymptotic except for one non-asymptotic estimation error bound for unit rank matrices $\C^*_k$ in \cite{mishra2017}.
However, the non-asymptotic error bound in \cite{mishra2017} does not admit an explicit form and is given in a recursive fashion, where the error bound for the $k$th unit rank matrix $\C^*_k$ is bounded
by the sum of estimation errors for the first $k-1$ unit rank matrices and four additional terms.
Of these four terms, one  involves the Frobenius norm of true $\C^*_k$ and another one measures the size of the left-over signal in the model.

The following proposition shows that our bounds in Theorem \ref{im:th} can be further improved when the response matrix $\Y$ is fully observed.

\begin{proposition}
	[Estimation and prediction bounds with full data]\label{com:th}
	Suppose that Condition \ref{cond1}--\ref{cond4} holds and the sparsity level $s\geq \max\limits_{1\leq k\leq r^*}s_k$ with $s_k=\norm{\u_k^*}_0$. Choose $\lambda_k = 4\tilde{c}\gamma_{d}^{-1}\widetilde{B}_{n}$ with $\tilde{c}$ some positive constant and
	$\widetilde{B}_{n}=\sigma\left(\frac{1}{\sqrt{n}}+\frac{1}{\sqrt{q}}\right)$. Then with probability at least $1- 2e^{-(\sqrt{n}+\sqrt{q})^2}$, the following inequalities
	\begin{align*}
		\norm{\widehat{\u}_k - \u_k^*}_2 \lesssim (\gamma_{d}\rho_l)^{-1}
		\sqrt{s_k}\widetilde{B}_n,\quad\quad %\label{com: uk-bound}
		 & \norm{\widehat{\v}_k - \v_k^*}_2 \lesssim \gamma_{d}^{-1} \widetilde{B}_n,                                                    %\label{com: vk-bound}
		\\
		q^{-1/2}\abs{\widehat{d}_k - d_k^*} \lesssim  \widetilde{B}_n,\quad\quad %\label{com: dk-bound}
		 & q^{-1/2}\norm{\widehat{\C}_{k}-\C^*_{k}}_{F} \lesssim (\gamma_{d}\rho_l)^{-1}
		\sqrt{s_k}\widetilde{B}_n, %\label{com: Ck-bound}
		\\
		\frac{\norm{\X(\widehat{\u}_k - \u_k^*)}_2}{\sqrt{n}} \lesssim \frac{\sqrt{s_k}\widetilde{B}_n}{\gamma_{d}\sqrt{\rho_l}},\quad\quad %\label{com: Xuk-bound}
		 & \frac{\norm{\X(\widehat{\C}_{k}-\C^*_{k})}_{F}}{\sqrt{nq}} \lesssim \frac{\sqrt{s_k}\widetilde{B}_n}{\gamma_{d}\sqrt{\rho_l}} %\label{com: XCk-bound}
	\end{align*}
	hold uniformly over $k=1, \cdots, r^*$.
\end{proposition}

To see the difference between the bounds of PEER with incomplete outcomes and with %complete outcomes 
full data, write $\alpha=m/(nq)$. Then $0<\alpha\leq 1$ and $1-\alpha$ is the missing rate of the response matrix $\Y$. It follows from the definition of $B_n$ in \eqref{eq: Bn-def}
that the second term in \eqref{eq: Bn-def} will dominate the first term if $n\geq 2$ by noting that $\nu\geq 1$, $\mu\geq 1$ and $m\leq nq$. Without loss of generality, assume $n\geq 2$. Thus we have $B_n=O\left(\sqrt{m^{-1}r(n \vee q)\log(n+q)}\right)=O\left(\sqrt{1/(n\wedge q)}\cdot \sqrt{r\log(n+q)/\alpha}\,\right)$ since $nq=(n\wedge q)(n \vee q)$. Note that the term $\sqrt{1/(n\wedge q)}$ has the same order as $\widetilde{B}_n$.
Therefore, the factor $\sqrt{r\log(n+q)/\alpha}$ reflects
the price we pay in dealing with incomplete outcomes, implying that smaller $\alpha$ leads to larger estimation and prediction errors.  It is reasonable since smaller $\alpha$ means that more entries of $\Y$ cannot be observed and thus corresponds
to more challenging case. This has also been observed in our simulations.

When the response matrix $\Y$ is fully observed, \cite{zheng2019} showed that the error bounds for $\u_k^*$, $\C^*_{k}$, $\X\u_k^*$, and $\X\C_k^*$ are all in the
same order of $O\left(\sqrt{\frac{s}{n}\log(pq/\delta)}\,\right)$ with some constant $\delta\in(0, 1)$ while our error bounds for those are in the order of $O\left(\sqrt{s_k}(\frac{1}{\sqrt{n}}+\frac{1}{\sqrt{q}})\right)$, which is  $O\left(\sqrt{s}(\frac{1}{\sqrt{n}}+\frac{1}{\sqrt{q}})\right)$ under the same assumption $\max\limits_{1\leq k\leq r^*}s_k\leq s$ as used in \cite{zheng2019}.
The factor $\log(pq)$ in the error bounds
of \cite{zheng2019} can become large and may not be negligible when either the number of predictors $p$ or the number of responses $q$ grows rapidly with sample size $n$. This indicates that our procedure PEER is preferred for big data applications.

The results in Theorem \ref{im:th} indicate that the regression coefficient matrix $\C^*$ can be accurately recovered and prediction error $(nq)^{-1/2}\norm{\X(\widehat{\C}-\C^*)}_{F}$ can be controlled if
the true rank $r^{*}$ is correctly identified. In particular, we use the following method to tune the true rank and have established its consistency.

\begin{theorem}[Consistency of rank recovery]\label{th3}
	Suppose Conditions \ref{cond1}, \ref{cond2}, \ref{cond3} and \ref{cond5} hold and $m^{-1}r(n\vee q)(\log ^2 n)\log(n+q)=O(1)$. We estimate the true rank $r^*$ by
	\begin{equation}\label{eq: rank-hat}
		\widehat{r}=\arg\max_{k}\{1\leq k\leq r: (nq)^{-1/2}(\widetilde{d}_k-\widetilde{d}_{k+1})>\tau_n\},
	\end{equation}
	where $\tau_n=(\log n)^{-1}\log\log n$. Then for sufficiently largely $n$, we have $\widehat{r}=r^*$ with probability at least $1-(n+q)^{-1}$.
\end{theorem}

Since the response matrix $\Y$ may not be fully observed,
the GIC-type \citep{fan2013} information criterion proposed in \cite{zheng2019} cannot be used to tune the true rank here. In addition, compared with tuning the true rank via cross validation, our method in \eqref{eq: rank-hat} enjoys much lower computational cost.

\section{Simulation Studies}\label{sec: sim}

In this section, we evaluate the finite-sample performance of the proposed approach PEER through
two simulation studies.
The main difference between these two studies lies in right singular vectors $\v_k^*$'s. The right singular vectors $\v_k^*$'s are sparse in the second study but not necessarily sparse in the first study.

\subsection{Study 1}\label{Study1}

We first state some model setups and simulation settings used in our numerical studies. For each $k=1, \cdots, r^*$, the sparse left singular
vector $\u_k^*$ is generated
with $\u_k^* = \bar{\u}_k / \norm{\bar{\u}_k}_2$, where $\bar{\u}_k =\left(\text{rep}(0, sk-s),\text{unif}(\mathcal{Q}_u, s),\text{rep}(0, p-sk)\right)\trans$. Here $\text{unif}(\mathcal{Q}, s)$ denotes a vector of length $s$ whose entries are i.i.d. uniformly distributed on set $\mathcal{Q}$ and $\text{rep}(a, b)$ represents a vector of length $b$ whose entries are all equal to $a$. Then we generate a matrix $\bar{\V}=\left(\bar{\v}_1,\cdots,\bar{\v}_{r^*}\right)\in\R^{q\times r^*}$ where $\bar{\v}_k = \text{unif}(\mathcal{Q}_v, q)$ for $k=1, \cdots, r^*$ and compute the QR decomposition $\bar{\V}=\Q{{\bf R}}$ where $\Q\in\R^{q\times r^*}$ with $\Q\trans\Q=\I_{r^*}$ and ${{\bf R}}\in\R^{r^*\times r^*}$ is a triangular matrix. Take the $k$th column of $\Q$ as $\v_k^*$ for $k=1,\cdots,r^*$.
For the singular values,
let $d_{k}^*=5+5(r^*-k+1)$ for $k=1, \cdots, r^*$.
The true coefficient matrix $\C^*$ is constructed as
$\C^* = \U^*\D^*\V\strans$ with $\U^* = \left(\u_1^*,\dots,\u_{r^*}^*\right)$, $\V^* = \left(\v_1^*,\dots,\v_{r^*}^*\right)$ and $\D^* = \text{diag}\{d_1^*,\dots,d_{r^*}^*\}$.

We use the similar procedure described in \citet{mishra2017} to generate the design matrix $\X$. More specifically,
let $\x \sim N(\0, \bGamma)$, where $\bGamma = (\gamma_{ij})_{p\times p}$ with $\gamma_{ij} = 0.5^{\abs{i-j}}$. Given $\U^* = \left(\u_1^*,\dots,\u_{r^*}^*\right)$, we can find $\U_{\bot}^* \in \mathbb{R}^{p\times (p-r^*)}$ such that ${\bf P} = (\U^*, \U_{\bot}^*)\in\mathbb{R}^{p\times p}$ and $\text{rank}({\bf P})=p$. Denote $\x_1=\U\strans\x$ and $\x_2 = \U_{\bot}\strans\x$. We first generate a matrix $\X_1\in\R^{n\times r^*}$ whose entries are from $N(\0,\I_{r^*})$ and then we generate $\X_2\in \mathbb{R}^{n\times (p-r^*)}$ by drawing $n$ random samples
from the conditional distribution of $\x_2$ given $\x_1$. The design matrix is then set as $\X=\left(\X_1,\X_2\right){\bf P}^{-1}$.

The entries of the error matrix $\E$ are generated as i.i.d. samples from $N(0,\sigma^2)$. Here $\sigma^2$ is chosen such that the signal to noise ratio (SNR), defined as $\text{SNR} = \norm{d_{r^*}^*\X\u_{r^*}^*\v_{r^*}\strans}_{F}/\norm{\E}_F$, is equal to a given value. %We consider $\text{SNR}\in\{0.25,\, 0.5\}$.  
Finally, the response matrix $\Y$ is generated by $\Y = \X\C^* + \E$. To obtain the data with incomplete outcomes, we randomly remove some entries in $\Y$ such that $(1-\alpha)\times 100\%$
percentage of entries in %the response matrix 
$\Y$ are unobserved. We consider four different values for the missing rate $1-\alpha$: $0$, $0.05$, $0.1$, and $0.15$. The case of missing rate $0$ means that $\Y$ is fully observed.

We consider four different settings: $(p, \text{SNR})=(200, 0.25)$, $(400, 0.25)$, $(200, 0.5)$, and $(400, 0.5)$. In all settings, we take $r^*=3$, $s=4$, %$n=100$, 
$q=100$, and $\mathcal{Q}_{u}=\left\{1,-1\right\}$, $\mathcal{Q}_{v} = \left[-1,-0.3\right]\cup\left[0.3,1\right]$. In all simulations, the regularization parameter $\lambda_k$ in \eqref{eq: uk-optim} is tuned by the following  GIC-type \citep{fan2013} information criterion
\begin{align*}
	\widehat{\lambda}_k=\arg\min_{\lambda_k}\left\{ \log(n^{-1}\norm{\sqrt{n}\,\widetilde{\z}_k - \X\widehat{\u}_k(\lambda_k)}_2^2) + n^{-1}\norm{\widehat{\u}_k(\lambda_k)}_0(\log p)(\log\log n)\right\}.
\end{align*}
%where $\widehat{\u}_k=\widehat{\u}_k(\lambda_k)$.
The experiment under each setting is repeated 200 times.

We first examine the impact of the missing rate on the performance of our method PEER through the following four measures. The estimation accuracy is measured by
$\text{Er}(\widehat{\C}) = \norm{\widehat{\C}-\C^*}_{F}^2/(pq)$ while
the prediction performance is measured by  $\text{Er}(\X\widehat{\C}) = \norm{\X(\widehat{\C} - \C^*)}_{F}^2/(nq)$. The variable selection performance is characterized by the false positive rate (FPR) and false negative rate (FNR) in recovering the sparsity patterns of the left singular vectors $\u_k$, where $\text{FPR} = \text{FP}/(\text{TN}+\text{FP})$ and $\text{FNR} = \text{FN}/(\text{TP}+\text{FN})$ with TP, FP, TN and FN being the numbers of true nonzeros, false nonzeros, true zeros, and false zeros of $\left\{\widehat{\u}_1,\dots,\widehat{\u}_{r^*}\right\}$, respectively.

\begin{figure}[ht!]
	\centering
	\includegraphics[scale=0.65]{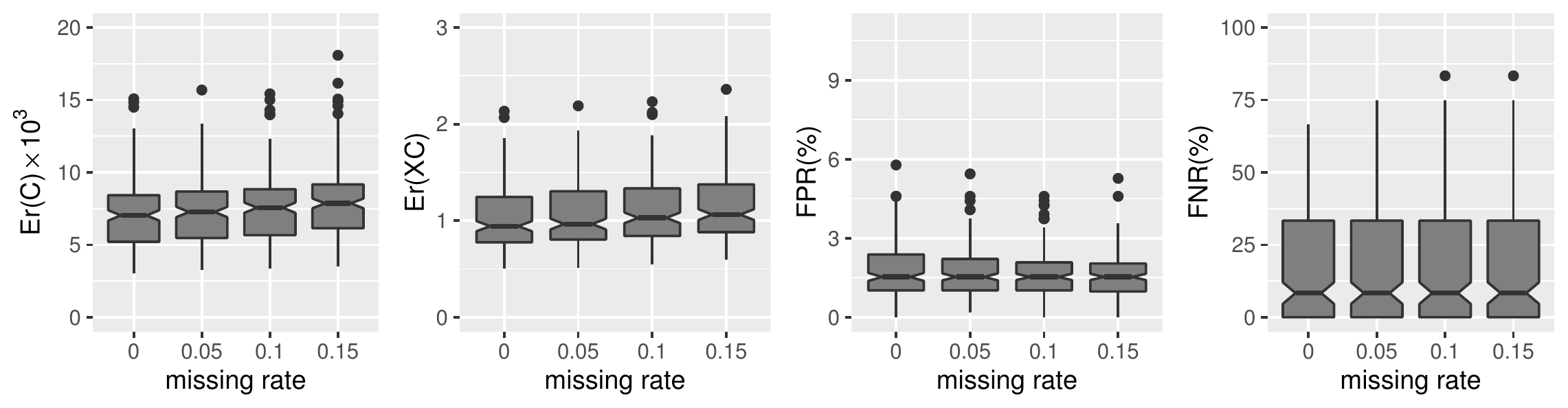}
	\includegraphics[scale=0.65]{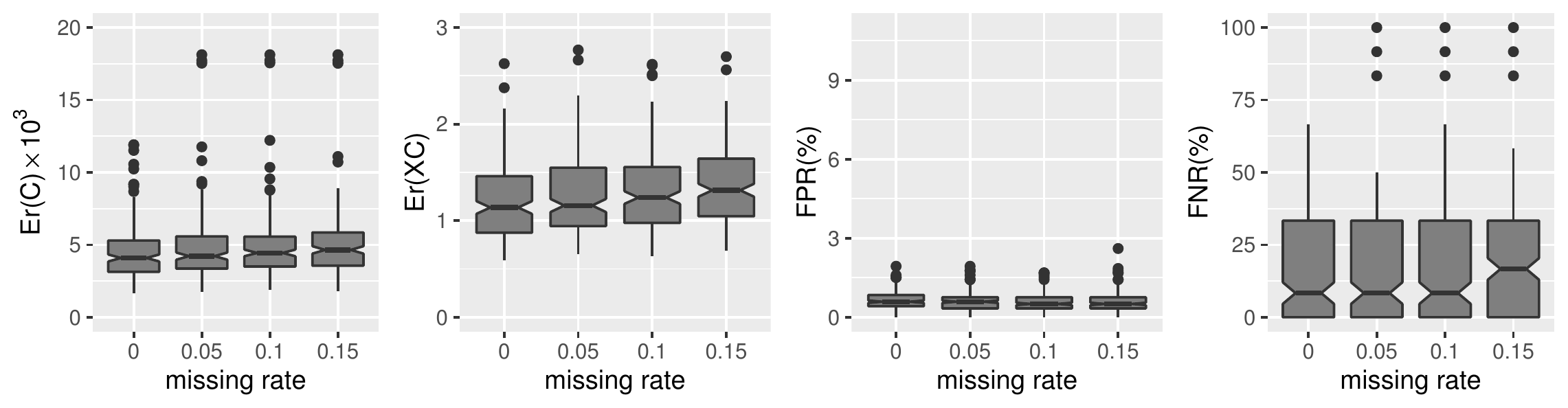}
	\includegraphics[scale=0.65]{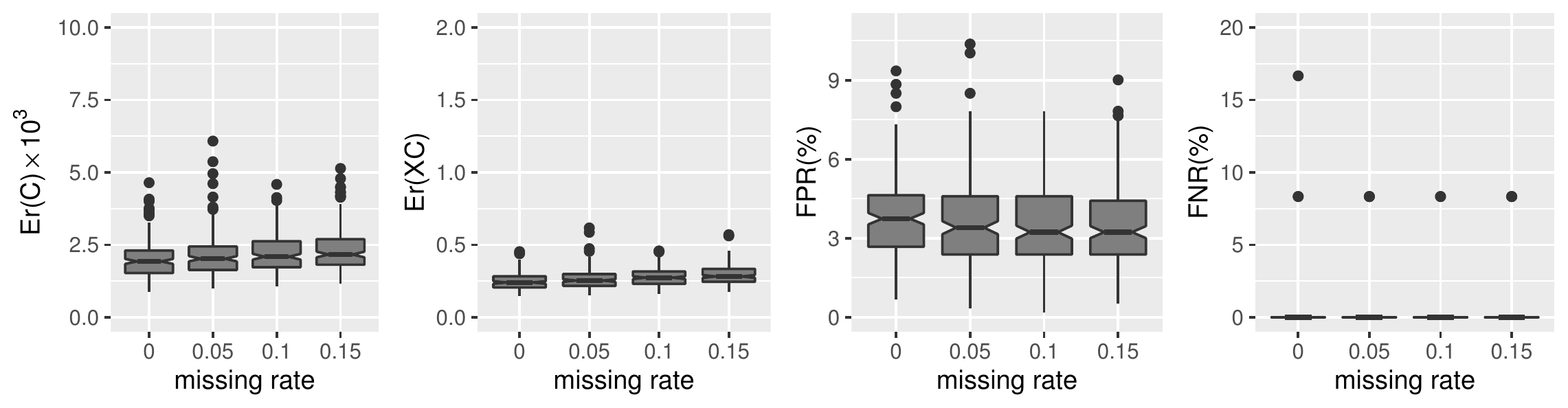}
	\includegraphics[scale=0.65]{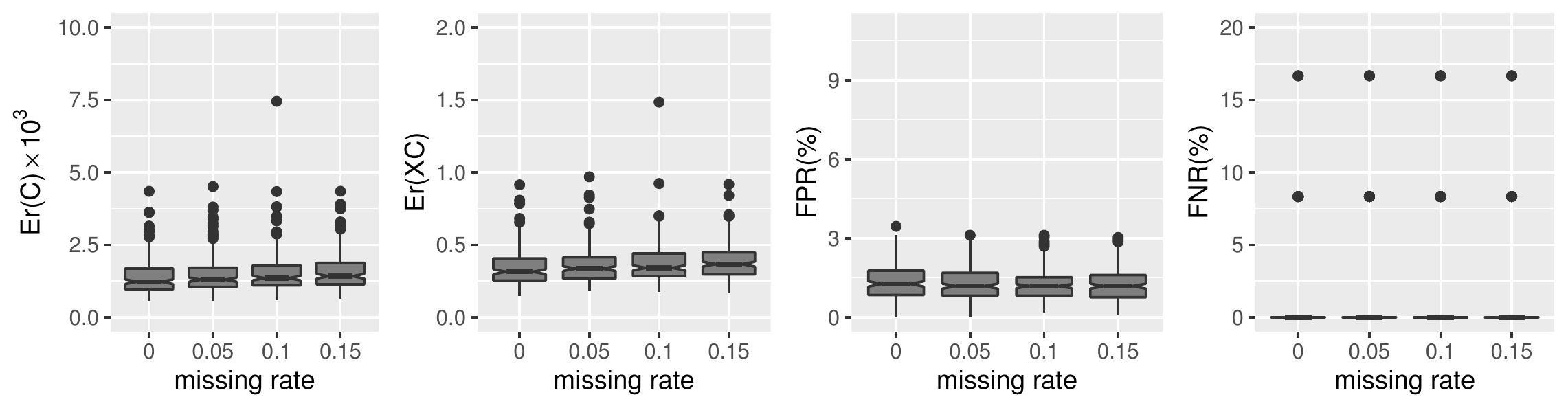}
	\caption{Impact of the missing rate on the performance of our method PEER when sample size $n=100$.
		First row: $(p, \text{SNR})=(200, 0.25)$;  second row: $(p, \text{SNR})=(400, 0.25)$;
		third row: $(p, \text{SNR})=(200, 0.5)$; fourth row: $(p, \text{SNR})=(400, 0.5)$.
	}\label{fig: miss-rev}
\end{figure}

Figure \ref{fig: miss-rev} presents the boxplots of the estimation error,
prediction error, false positive rate and false negative rate for our method PEER when sample size $n=100$. It shows that the estimation error %$\text{Er}(\widehat{\C})$ 
and prediction error %$\text{Er}(\X\widehat{\C})$ 
increase with the missing rate, %as the missing rate becomes larger, 
which is consistent with our theory in Theorem \ref{im:th}. It is also clear that the variable selection performance of PEER is
robust to the missing rate since both false positive rates and false negative rates are stabilized.  %The computational cost will increase if the missing rate is large. 

Next we compare our proposed method PEER with other competing approaches, including sequential factor extraction via co-sparse unit-rank estimation (SeCURE) \citep{mishra2017}, and mixed-response reduced-rank regression (mRRR) \citep{luo2018}. We do not include the sequential estimation with eigen-decomposition (SEED) \citep{zheng2019} in our comparison because SEED is not applicable for incomplete outcomes. In addition to the four measures $\text{Er}(\X\widehat{\C})$, $\text{Er}(\X\widehat{\C})$, FPR, and FNR, we also compare these methods in terms of the computational cost, which is evaluated by average running time (in seconds) of 200 replicates and denoted by Time.
We employ the R packages \code{rrpack} \citep{rrpack}
and \code{secure} \citep{secure} to implement mRRR and SeCURE, respectively. We use 5-fold cross validation to select the rank of
mRRR and the threshold $\tau_n=(\log\log n) /\log n$ to tune
the rank of PEER. The maximum rank of SeCURE is set as $4$.
We consider two different sample sizes $n=100$ and $n=200$.

\begin{table}[htbp]
	\centering
	\resizebox{0.9\columnwidth}{!}{
		\begin{tabular}{cccccc}
			\toprule
			Method & $\text{Er}(\widehat{\C})\times 10^3$ & $\text{Er}(\X\widehat{\C})$ & FPR (\%)      & FNR (\%)      & Time (s)         \\
			\midrule\multicolumn{6}{c}{$p=200$, $\text{SNR}=0.25$}                                                                         \\
			mRRR   & 21.65 (1.41)                         & 1.20 (0.20)                 & 96.83 (9.80)  & 3.17 (9.80)   & 366.91 (116.29)  \\
			SeCURE & 36.15 (0.61)                         & 7.23 (0.69)                 & 0.00 (0.01)   & 99.00 (4.77)  & 3.10 (2.12)      \\
			PEER   & 12.81 (9.32)                         & 2.19 (1.86)                 & 1.38 (0.97)   & 35.63 (19.98) & 1.06 (0.12)      \\

			\midrule\multicolumn{6}{c}{$p=400$, $\text{SNR}=0.25$}                                                                         \\
			mRRR   & 14.40 (0.42)                         & 1.33 (0.27)                 & 91.00 (14.84) & 9.00 (14.84)  & 765.45 (267.19)  \\
			SeCURE & 18.10 (0.19)                         & 7.17 (0.67)                 & 0.00 (0.00)   & 99.25 (3.36)  & 10.81 (6.02)     \\
			PEER   & 7.28 (4.61)                          & 2.43 (1.83)                 & 0.49 (0.36)   & 38.67 (19.27) & 1.07 (0.12)      \\

			\midrule\multicolumn{6}{c}{$p=200$, $\text{SNR}=0.5$}                                                                          \\
			mRRR   & 18.42 (1.28)                         & 0.29 (0.07)                 & 99.83 (2.36)  & 0.17 (2.36)   & 654.67 (153.86)  \\
			SeCURE & 34.80 (1.48)                         & 6.74 (0.70)                 & 0.02 (0.13)   & 90.50 (9.56)  & 2.90 (1.68)      \\
			PEER   & 2.20 (0.78)                          & 0.27 (0.07)                 & 3.41 (1.42)   & 0.29 (2.27)   & 1.05 (0.12)      \\

			\midrule\multicolumn{6}{c}{$p=400$, $\text{SNR}=0.5$}                                                                          \\
			mRRR   & 13.61 (0.43)                         & 0.32 (0.16)                 & 98.50 (6.93)  & 1.50 (6.93)   & 1323.48 (272.69) \\
			SeCURE & 17.30 (0.82)                         & 6.68 (0.72)                 & 0.01 (0.06)   & 89.63 (10.55) & 10.57 (5.20)     \\
			PEER   & 1.65 (0.74)                          & 0.39 (0.15)                 & 1.27 (0.60)   & 1.25 (4.40)   & 1.07 (0.12)      \\
			\bottomrule
		\end{tabular}
	}
	\caption{The means and standard errors (in parentheses) of various performance measures for
		different approaches in Study 1 with $n=100$ and missing rate $0.1$.} \label{tab:miss:1-rev-n-100}
\end{table}

Table \ref{tab:miss:1-rev-n-100} reports the comparison results with different $p$ and SNR when sample size $n=100$ and $10\%$ of entries of $\Y$ are not observed. It can be seen that PEER has better performance than mRRR and SeCURE. Firstly, we can see that PEER has smaller estimation error $\text{Er}(\widehat{\C})$ than mRRR and SeCURE across all settings. In terms of
prediction error $\text{Er}(\X\widehat{\C})$, PEER is
superior to SeCURE under all settings and comparable to mRRR (especially when SNR=0.5).
Secondly, compared to PEER, other approaches result in either larger false positive rates or larger false negative rates for all settings.
Thirdly, PEER is much faster than other approaches.
We can
see that PEER can achieve a speed up of about 3-10 times in runtime compared with SeCURE and be more than 1000 times faster than mRRR.

\begin{table}[htbp]
	\centering
	\resizebox{0.9\textwidth}{!}{
		\begin{tabular}{cccccc}
			\toprule
			Method & $\text{Er}(\widehat{\C})\times 10^3$ & $\text{Er}(\X\widehat{\C})$ & FPR (\%)      & FNR (\%)      & Time (s)         \\

			\midrule\multicolumn{6}{c}{$p=200$, $\text{SNR}=0.25$}                                                                         \\

			mRRR   & 11.49 (1.12)                         & 0.71 (0.07)                 & 100.00 (0.00) & 0.00 (0.00)   & 333.99 (41.32)   \\
			SeCURE & 35.74 (1.17)                         & 7.09 (0.58)                 & 0.02 (0.12)   & 96.58 (7.76)  & 5.83 (3.70)      \\
			PEER   & 3.30 (1.26)                          & 0.51 (0.21)                 & 2.21 (1.00)   & 2.58 (8.82)   & 2.08 (0.22)      \\

			\midrule\multicolumn{6}{c}{$p=400$, $\text{SNR}=0.25$}                                                                         \\

			mRRR   & 10.36 (0.48)                         & 0.80 (0.09)                 & 99.83 (2.36)  & 0.17 (2.36)   & 556.37 (145.22)  \\
			SeCURE & 17.45 (0.79)                         & 6.78 (0.56)                 & 0.01 (0.05)   & 91.04 (10.37) & 16.69 (9.56)     \\
			PEER   & 1.90 (0.65)                          & 0.56 (0.21)                 & 0.85 (0.43)   & 2.71 (9.07)   & 2.01 (0.23)      \\

			\midrule\multicolumn{6}{c}{$p=200$, $\text{SNR}=0.5$}                                                                          \\

			mRRR   & 6.51 (0.82)                          & 0.19 (0.02)                 & 100.00 (0.00) & 0.00 (0.00)   & 497.60 (200.45)  \\
			SeCURE & 34.04 (1.77)                         & 6.58 (0.65)                 & 0.25 (0.48)   & 89.00 (11.63) & 6.73 (3.72)      \\
			PEER   & 0.90 (0.20)                          & 0.13 (0.02)                 & 3.71 (1.36)   & 0.00 (0.00)   & 1.97 (0.21)      \\

			\midrule\multicolumn{6}{c}{$p=400$, $\text{SNR}=0.5$}                                                                          \\

			mRRR   & 9.05 (0.53)                          & 0.20 (0.02)                 & 100.00 (0.00) & 0.00 (0.00)   & 1132.99 (352.77) \\
			SeCURE & 16.47 (0.67)                         & 6.16 (0.54)                 & 0.07 (0.13)   & 79.08 (15.27) & 18.37 (10.02)    \\
			PEER   & 0.55 (0.16)                          & 0.15 (0.03)                 & 1.56 (0.68)   & 0.00 (0.00)   & 1.96 (0.19)      \\
			\bottomrule
		\end{tabular}%
	}
	\caption{The means and standard errors (in parentheses) of various performance measures for
		different approaches in Study 1 with $n=200$ and missing rate $0.1$.} \label{tab:miss:1-rev-n-200}
\end{table}

The comparison results with sample size $n=200$ are reported in Table \ref{tab:miss:1-rev-n-200}, which shows the similar results as Table \ref{tab:miss:1-rev-n-100}. In particular, we can see that PEER has smaller estimation error $\text{Er}(\widehat{\C})$ and prediction error $\text{Er}(\X\widehat{\C})$ than other approaches across all settings. By comparing the results for the same approach in Tables \ref{tab:miss:1-rev-n-100} and \ref{tab:miss:1-rev-n-200}, it is clear that larger sample size can lead to smaller estimation error $\text{Er}(\widehat{\C})$ and prediction error $\text{Er}(\X\widehat{\C})$ for both mRRR and PEER.

We should remark that in this study the true right singular vectors $\v_k^*$'s are not necessarily sparse. However, SeCURE assumes right singular vectors $\v_k^*$'s to be sparse. This may be the reason why SeCURE has larger estimation error $\text{Er}(\widehat{\C})$ and prediction error $\text{Er}(\X\widehat{\C})$ in this study.
Thus,  we consider another study where the true $\v_k^*$'s are sparse in next subsection to check this.

\subsection{Study 2}\label{Study2}

Following \cite{mishra2017}, we
generate sparse right singular vectors $\v_k^*$'s in this study. %, which favors SeCURE. 
To be specific, we generate $\u_k^* = \bar{\u}_k/\norm{\bar{\u}_k}_2$ and $\v_k^*= \bar{\v}_k$,
where $\bar{\u}_k = (\text{rep}(0,ks_u-s_u), \text{unif}(\mathcal{Q}_u,s_u), \text{rep}(0,p-ks_u))$ and $\bar{\v}_k = (\text{rep}(0,ks_v-s_v), \text{unif}(\mathcal{Q}_v,s_v), \text{rep}(0,q-ks_v))$ with $s_u=4$ and $s_v=5$, meaning that $\norm{\u_k^*}_0=4$ and $\norm{\v_k^*}_0=5$ for each $k=1,\dots,r^*$. We take $n=200$ and consider four different settings $(p, \text{SNR})=(200, 0.5)$, $(400, 0.5)$, $(200, 1)$, and $(400, 1)$. The other settings are the same as those in Study 1. Then we follow the same procedure as in Study 1 to generate the true coefficient matrix $\C^*$, the design matrix $\X$ and the response matrix $\Y$.

\begin{table}[ht!]
	\centering
	\resizebox{0.9\textwidth}{!}{
		\begin{tabular}{cccccc}
			\toprule
			Method & $\text{Er}(\widehat{\C})\times 10^3$ & $\text{Er}(\X\widehat{\C})$ & FPR (\%)    & FNR (\%)      & Time (s)       \\
			\midrule\multicolumn{6}{c}{$p=200$, $\text{SNR}=0.5$}                                                                      \\
			SeCURE & 6.32 (8.54)                          & 0.90 (1.58)                 & 9.17 (7.31) & 19.54 (28.03) & 15.11 (4.64)   \\
			PEER   & 0.81 (0.18)                          & 0.12 (0.02)                 & 2.41 (1.30) & 0.00  (0.00)  & 2.36  (0.34)   \\

			\midrule\multicolumn{6}{c}{$p=400$, $\text{SNR}=0.5$}                                                                      \\
			SeCURE & 0.04 (0.04)                          & 0.01 (0.01)                 & 0.05 (0.16) & 0.00 (0.00)   & 92.58 (18.31)  \\
			PEER   & 0.49 (0.13)                          & 0.14 (0.03)                 & 1.07 (0.55) & 0.00 (0.00)   & 2.38 (0.37)    \\

			\midrule\multicolumn{6}{c}{$p=200$, $\text{SNR}=1$}                                                                        \\
			SeCURE & 2.18 (5.37)                          & 0.29 (0.93)                 & 5.48 (6.21) & 8.83 (21.15)  & 22.43 (4.55)   \\
			PEER   & 0.24 (0.06)                          & 0.03 (0.01)                 & 3.87 (1.37) & 0.00 (0.00)   & 2.38 (0.40)    \\

			\midrule\multicolumn{6}{c}{$p=400$, $\text{SNR}=1$}                                                                        \\
			SeCURE & 0.03 (0.04)                          & 0.01 (0.01)                 & 0.15 (0.28) & 0.00 (0.00)   & 108.23 (19.46) \\
			PEER   & 0.15 (0.04)                          & 0.04 (0.01)                 & 1.64 (0.64) & 0.00 (0.00)   & 2.38 (0.35)    \\
			\bottomrule
		\end{tabular}
	}
	\caption{The means and standard errors (in parentheses) of various performance measures for
		SeCURE and PEER in Study 2 with missing rate $0.1$.}\label{tab-study2}
\end{table}

Table \ref{tab-study2} records the simulation results for Study 2 under different settings. As expected, SeCURE has good performance when the true right singular vectors $\v_k^*$'s are also sparse. It is also interesting to see that PEER still has good performance in this study. This is not surprising because PEER does not require the sparsity assumption on $\v_k^*$ and can deal with both sparse $\v_k^*$ and non-sparse $\v_k^*$. Table \ref{tab-study2} also shows that the performance of both methods improves when SNR becomes higher.

\section{Yeast cell cycle data analysis}\label{sec:yeast}

In this section, we apply the proposed method to a multivariate Yeast cell cycle data,
%We further evaluate the performance of our proposed
%method PEER by a multivariate Yeast cell cycle data analysis, 
in which our goal is to identify the association between transcription factors (TFs) and RNA transcript levels within the Eukaryotic cell cycle.  The dataset that we used includes the yeast cell cycle data originally collected by \cite{spellman1998comprehensive} and the chromatin immunoprecipitation (ChIP) data in \cite{Lee2002}. The yeast cell cycle data in  \cite{spellman1998comprehensive} consist of RNA levels measured
every $7$ minutes for $119$ minutes with a total of $18$ time points covering two cell cycle of
$6178$ genes while the ChIP data in \cite{Lee2002} contain
complete binding information of %only $1790$ genes
only a subset of these 6178 genes of size 1790 for a total of 113 TFs. In addition, the RNA levels corresponding to these 1790 genes in the yeast cell cycle data contain about $2\%$ missing values.

Here, we use these $1790$ genes with RNA transcript levels at $18$ time points and
binding information of 113 TFs to examine the
association between the TFs and the RNA transcript levels. Thus, our response matrix $\Y$ is a $1709\times 18$ matrix, recording RNA levels of
$1709$ genes at $18$ time points, and there are about $2\%$ missing values in $\Y$. Our design matrix $\X$ is a $1709\times 113$ matrix, corresponding to complete binding information of these $1790$ genes for a total of 113 TFs.
In other words, our data set has sample size $n=1709$, number of covariates $p=113$, and number of responses $q=18$.

\begin{table}[!htb]
	\centering
	\resizebox{0.9\columnwidth}{!}{
		\begin{tabular}{c|c|c}
			\toprule
			 & PEER                       & SeCURE \\
			\hline
			1th factor
			 & \begin{tabular}{ccccc}
				\textbf{ABF1} & CIN5           & FHL1          & HIR2          & MTH1 \\
				RGM1          & SKO1           & SOK2          & USV1          & YAP5 \\
				\textbf{REB1} & \textbf{STE12} & \textbf{NDD1} & \textbf{FKH2} &
			\end{tabular}
			 & \begin{tabular}{c}
				\textbf{NDD1}
			\end{tabular}          \\
			\hline
			2th factor
			 & \begin{tabular}{cccc}
				FHL1          & GAT3           & HSF1          & \textbf{MBP1} \\
				\textbf{NDD1} & \textbf{STE12} & \textbf{SWI4} & YAP5
			\end{tabular}
			 & \begin{tabular}{ccccc}
				\textbf{ACE2} & CUP9          & DAL81 & \textbf{STB1}  & \textbf{SWI4} \\
				\textbf{NDD1} & PDR1          & RTG1  & \textbf{MET31} & \textbf{SKN7} \\
				\textbf{SWI5} & \textbf{SWI6} & USV1  & \textbf{MBP1}  & GAT3          \\
				HIR1          & HIR2          & SKO1  & \textbf{STE12} & SIP4          \\
				DIG1
			\end{tabular}          \\
			\hline
			3th factor
			 & \begin{tabular}{ccccc}
				DAL82         & HSF1          & INO4 & SFL1          & SOK2 \\
				\textbf{MCM1} & \textbf{SWI5} & RAP1 & \textbf{SWI4} &
			\end{tabular}
			 & \begin{tabular}{c}
				\textbf{SWI5}
			\end{tabular}          \\
			\hline
			4th factor
			 & \begin{tabular}{cccc}
				\textbf{ACE2} & CIN5          & FHL1 & \textbf{STE12} \\
				\textbf{FKH2} & \textbf{NDD1} & RFX1 & \textbf{SWI4}
			\end{tabular}
			 & \begin{tabular}{cccc}
				\textbf{ACE2} & HIR1           & \textbf{MBP1} & \textbf{SWI4} \\
				\textbf{NDD1} & \textbf{STE12} & STP1          & YAP5
			\end{tabular}          \\
			\bottomrule
		\end{tabular}
	}
	\caption{TFs selected by PEER and SeCURE where
		those TFs in bold are confirmed to related to the cell cycle regulation in \cite{wang2007}.}
	\label{tab:yeast:factor}
\end{table}

The same dataset has been analyzed in \cite{mishra2017} and can be
accessed
in the R package \code{secure} \citep{secure}. Following \cite{mishra2017}, we consider the first four latent factors. Table \ref{tab:yeast:factor} presents
the variable selection results for these four latent factors, where TFs in bold have been %experimentally 
confirmed to be related to the cell cycle regulation in \cite{wang2007}. Our approach PEER selected 26 distinct TFs in total, of which 10 are among the confirmed TFs. SeCURE selected 23 distinct TFs with  10 confirmed TFs.  Of these confirmed TFs, 6 (ACE2, MBP1, NDD1, STE12, SWI4, SWI5) were selected by both methods. The 4 confirmed TFs (ABF1, FKH2, MCM1, REB1) were selected by PEER but missed by SeCURE.
However, these four TFs missed by SeCURE can be also important in cell cycle. For example, ABF1 and FKH2 play an important role in DNA/RNA/protein biosynthesis, and REB1 works for environmental response in cell cycle \citep{Lee2002}. We remark that our method PEER also missed 4 confirmed TFs (MET31, SKN7, STB1, SWI6), which were selected by SeCURE.  Therefore, our method can be used to uncover genes that are important but missed by SeCURE.  For variable selection purpose, we can use both methods in practice to avoid missing important features.

\section{Discussion}\label{sec:discussion}

In this paper, we have proposed a new and efficient approach PEER to achieve scalable and accurate estimation for large-scale multi-response regression with incomplete outcomes, where
both responses and predictors are possibly of high dimensions.
It has been shown through our theoretical properties and numerical studies that PEER achieves nice estimation and prediction accuracy.

Here we have focused on multi-response linear models with
incomplete outcomes, where all the responses are continuous variables.
In many applications, linear models can become restrictive. In addition, the responses can be categorical, counts, or functional in many real-world problems.
For more flexible modeling,
it is also of practical importance to extend the idea of PEER to more general model settings with incomplete outcomes, such as multi-response generalized linear models \citep{dette2013optimal}, partial linear multiplicative models \citep{zhang2018estimation}, multivariate varying coefficient models \citep{he2018dimensionality},
semiparametric quantile factor models \citep{ma2020estimation},
and multivariate functional responses \citep{liu2020multivariate}.
These possible extensions are beyond the
scope of the current paper and will be interesting topics for future research.

%
%\section*{Acknowledgments}
%
%Li's research was supported by 2020 individual Award (0358220) from the Innovative Research and Creative Activities Grant at California State University, Fullerton. Zheng's research was
%supported by National Natural Science Foundation of China (Grants 72071187, 11671374, 71731010, and 71921001) and Fundamental Research Funds for the Central Universities (Grants WK3470000017 and WK2040000027). The authors sincerely thank the Co-Editor, Associate Editor, and anonymous referees for their valuable and constructive comments that helped improve the article substantially.
%

\appendix
\section{Proofs of main results}\label{sec: appendix}

To ease the presentation, we further introduce some notation which will be used later. Let $\inner{\A, \B}$ be the Frobenius inner product of
two matrices $\A$ and $\B$. We also use $\inner{\a,\b}$ to denote the inner product of
two vectors $\a$ and $\b$. For each $k=1,\dots,r^*$, let $J_k=\{1\leq j\leq p:u_{k,j}^*\neq 0\}$ be the support of $\u_k^*$ where $u_{k,j}^*$ is the $j$th element of $\u_k^*$.
In addition, define $\widehat{\bdelta}_k=\widehat{\u}_k - \u_k^*$, where $\widehat{\u}_k$ is the estimation of $\u_k^*$. Hereafter we use $c$ to denote a generic positive constant whose value may vary from place to place.

We first present two additional lemmas. These results can also be of independent interest. Lemmas \ref{incomp:lem2} and \ref{incomp:lem1} will be used in the proof of Theorem \ref{im:th} while lemma \ref{comp:lem1} will be used in the proof of Theorem \ref{com:th}. The proofs of
all lemmas are provided in \ref{App-B}.

\begin{lemma}\label{incomp:lem1}
	Assume that $\norm{\X\C^*}_{\max} \leq L$ and $m \geq\nu^{-1} (n\wedge q)\log^3(n+q)$. Then, under Conditions \ref{cond1} and \ref{cond3}, we have that $\norm{\widetilde{\bDelta}}_F/\sqrt{nq}\leq c B_n$ holds
	with probability at least $1-(n+q)^{-1}$, where $c$ is a positive constant,
	$\widetilde{\bDelta}=\widetilde{\Y} - \X\C^*$, and $\widetilde{\Y}$ and $B_n$ are defined in \eqref{opt:matrix-completion} and \eqref{eq: Bn-def}, respectively.
\end{lemma}

\begin{lemma}\label{comp:lem1}
	Under Conditions \ref{cond1} and \ref{cond2}, we have that, with probability at least $1 - 2e^{-(\sqrt{n}+\sqrt{q})^2}$,
	\begin{align*}
		 & \max\left\{\norm{\widetilde{\z}_k - n^{-1/2} \X\u_k^*}_2, \,\, \norm{\widetilde{\v}_k - \v_k^*}_2\right\}
		\lesssim \gamma_{d}^{-1}\widetilde{B}_n,                                                                     \\
		 & ~\abs{\widetilde{d}_k - \sqrt{n}d_k^*}/\sqrt{nq}
		\lesssim\widetilde{B}_n,
	\end{align*}
	hold uniformly over $k=1, \cdots, r^*$,
	where $\widetilde{B}_n=\sigma\left(\frac{1}{\sqrt{n}} + \frac{1}{\sqrt{q}}\right)$.
\end{lemma}

\subsection{Proof of Theorem \ref{im:th}}\label{App-im-th}

Define the event $\Omega =\left\{\norm{\widetilde{\bDelta}}_F/\sqrt{nq}\leq cB_n\right\}$.
Then it follows from Lemma \ref{incomp:lem1} that the event $\Omega$ holds with probability at least $1-(n+q)^{-1}$, that is, $P(\Omega)\geq 1-(n+q)^{-1}$.
Thus, to prove Theorem \ref{im:th}, it suffices to show that, conditional on the event $\Omega$, the following six inequalities
\begin{align*}
	n^{-1/2}\norm{\X(\widehat{\u}_k - \u_k^*)}_2
	 & \lesssim \frac{\sqrt{s_k}}{\gamma_{d}\sqrt{\rho_l}}B_n, \quad\,
	\norm{\widehat{\u}_k - \u_k^*}_2  \lesssim \frac{\sqrt{s_k}}{\rho_l\gamma_{d}}B_n, \\
	\norm{\widehat{\v}_k - \v_k^*}_2
	 & \lesssim \gamma_{d}^{-1}B_n,\quad\,
	\abs{\widehat{d}_k - d_k^*}/\sqrt{q} \lesssim B_n,                                 \\
	\frac{1}{\sqrt{nq}}\norm{\X(\widehat{\C}_{k}-\C^*_{k})}_{F}
	 & \lesssim \frac{\sqrt{s_k}}{\gamma_{d}\sqrt{\rho_l}}B_n,\quad\,
	\frac{1}{\sqrt{q}}\norm{\widehat{\C}_{k}-\C^*_{k}}_{F} \lesssim \frac{\sqrt{s_k}}{\rho_l\gamma_{d}}B_n
\end{align*}
hold uniformly over $k=1, \cdots, r^*$.

Hereafter our analysis will be conditional on the event $\Omega$.  By \eqref{eq: A13-Li},
we have that, conditional on the event $\Omega$, $	\norm{\widehat{\v}_k - \v_k^*}_2 \lesssim \gamma_{d}^{-1}B_n$ holds uniformly over $k=1, \cdots, r^*$.  Hence it remains to prove other five inequalities. To enhance readability, we split the proof into five parts.

	{\bf \underline{An upper bound for $n^{-1/2}\norm{\X(\widehat{\u}_k - \u_k^*)}_2$}}. Since $\widehat{\u}_k$ is the minimizer of of \eqref{eq: uk-optim}, we have
\begin{align*}
	n^{-1}\norm{n^{1/2}\widetilde{\z}_k - \X\widehat{\u}_k}_2^2 + \lambda_k\norm{\widehat{\u}_k}_1
	\leq n^{-1}\norm{n^{1/2}\widetilde{\z}_k - \X\u_k^*}_2^2 + \lambda_k\norm{\u_k^*}_1.
\end{align*}
Recall that $\widehat{\bdelta}_k = \widehat{\u}_k - \u_k^*$. Thus
$n^{1/2}\widetilde{\z}_k - \X\widehat{\u}_k = (n^{1/2}\widetilde{\z}_k - \X\u_k^*) - \X\widehat{\bdelta}_k$.
Substituting this into the above inequality, after some simple algebra, we obtain
\begin{align}\label{im:th:ineq1}
	n^{-1}\norm{\X\widehat{\bdelta}_k}_2^2 + \lambda_k\norm{\widehat{\u}_k}_1
	 & \leq
	2n^{-1}\inner{n^{1/2}\widetilde{\z}_k - \X\u_k^*, \X\widehat{\bdelta}_k} + \lambda_k\norm{\u_k^*}_1                                                                                        \nonumber \\
	 & = 2n^{-1}\inner{\X\trans(n^{1/2}\widetilde{\z}_k - \X\u_k^*), \widehat{\bdelta}_k} + \lambda_k\norm{\u_k^*}_1.
\end{align}
An application of the triangle inequality yields
\begin{align}\label{im:th:ineq2}
	n^{-1}\abs{\inner{\X\trans(n^{1/2}\widetilde{\z}_k - \X\u_k^*), \widehat{\bdelta}_k}} \leq
	n^{-1}\norm{\X\trans(n^{1/2}\widetilde{\z}_k - \X\u_k^*)}_{\infty}\norm{\widehat{\bdelta}_k}_1.
\end{align}
Write $\X=(\widetilde{\x}_1, \cdots, \widetilde{\x}_p)$, where $\widetilde{\x}_j$ is the $j$th column of $X$ for $j=1, \cdots, p$.
By the Cauchy-Schwarz inequality, we have
\begin{align*}
	     & n^{-1}\norm{\X\trans(n^{1/2}\widetilde{\z}_k - \X\u_k^*)}_{\infty}
	=\max_{1\leq j \leq p}\abs{n^{-1/2}\widetilde{\x}_j\trans(\widetilde{\z}_k - n^{-1/2}\X\u_k^*)}           \\
	\leq & \max_{1\leq j \leq p}n^{-1/2}\norm{\widetilde{\x}_j}_2\norm{\widetilde{\z}_k - n^{-1/2}\X\u_k^*}_2
	\leq \norm{\widetilde{\z}_k - n^{-1/2}\X\u_k^*}_2,
\end{align*}
where the last inequality holds since each column of $\X$ is rescaled to have an $\ell_2$-norm $n^{1/2}$.  Conditional on the event $\Omega$, this together with \eqref{eq: A12-Li} yields
\begin{align}\label{im:th:ineq3}
	n^{-1}\norm{\X\trans(n^{1/2}\widetilde{\z}_k - \X\u_k^*)}_{\infty} \leq c\gamma_{d}^{-1}B_n,
\end{align}
where $c$ is some positive constant. Taking
$\lambda_k = 4c\gamma_{d}^{-1}B_n$ in \eqref{im:th:ineq1} and combining it with \eqref{im:th:ineq2} and \eqref{im:th:ineq3} lead to
\begin{align}\label{im:th:ineq4}
	n^{-1}\norm{\X\widehat{\bdelta}_k}_2^2 + \lambda_k\norm{\widehat{\u}_k}_1 \leq
	\frac{\lambda_k}{2}\norm{\widehat{\bdelta}_k}_1 + \lambda_k\norm{\u_k^*}_1,
\end{align}

Let $\widehat{\u}_{J_k}$ and $\u^*_{J_k}$ be the subvectors of $\widehat{\u}_{k}$ and $\u^*_k$ formed by components in $J_k$, respectively. Similarly,
let $\widehat{\bdelta}_{J_k}$ and $\widehat{\bdelta}_{J_k^c}$ be
the subvectors of $\widehat{\bdelta}_k$ formed by components in $J_k$ and $J_k^c$, respectively. The inequality \eqref{im:th:ineq4} entails
\begin{align*}
	\norm{\widehat{\u}_k}_1 \leq
	\norm{\widehat{\bdelta}_k}_1/2 + \norm{\u_k^*}_1
	\leq (\norm{\widehat{\bdelta}_{J_k}}_1 + \norm{\widehat{\bdelta}_{J_k^c}}_1)/2+ \norm{\u_{J_k}^*}_1.
\end{align*}
Since $\norm{\widehat{\u}_k}_1 = \norm{\widehat{\u}_{J_k}}_1 + \norm{\widehat{\bdelta}_{J_k^c}}_1$, the above inequality yields
\begin{align}\label{im:th:ineq5}
	\norm{\widehat{\bdelta}_{J_k^c}}_1
	\leq \norm{\widehat{\bdelta}_{J_k}}_1+ 2\left(\norm{\u_{J_k}^*}_1-\norm{\widehat{\u}_{J_k}}_1\right)\leq 3\norm{\widehat{\bdelta}_{J_k}}_1,
\end{align}
where the last inequality follows from the reverse triangle inequality for $\|\cdot\|_1$.
Using \eqref{im:th:ineq4} and the reverse triangle inequality again gives
\begin{align*}
	n^{-1}\norm{\X\widehat{\bdelta}_k}_2^2
	\leq \lambda_k\norm{\widehat{\bdelta}_k}_1/2+
	\lambda_k\left(\norm{\u_k^*}_1-\norm{\widehat{\u}_k}_1\right)\leq 3\lambda_k \norm{\widehat{\bdelta}_k}_1/2.
\end{align*}
Thus, by \eqref{im:th:ineq5} and the Cauchy-Schwarz inequality, we have
\begin{align}\label{im:th:ineq6}
	n^{-1}\norm{\X\widehat{\bdelta}_k}_2^2
	\leq 3\lambda_k(\norm{\widehat{\bdelta}_{J_k}}_1 + \norm{\widehat{\bdelta}_{J_k^c}}_1)/2
	\leq 6\lambda_k\norm{\widehat{\bdelta}_{J_k}}_1
	\leq 6\lambda_k\sqrt{s_k}\norm{\widehat{\bdelta}_{J_k}}_2,
\end{align}
where $s_k=\norm{\u_k^*}_0$ is equal to the cardinality of the set $J_k$.
This, together with \eqref{im:th:ineq5} and Condition \ref{cond4}, yields
\begin{align*}
	\max(\norm{\widehat{\bdelta}_{J_k}}_2^2,~ \norm{\widehat{\bdelta}^{(1)}_{J_{k}^c}}_2^2)
	\leq
	\rho_l^{-1}n^{-1}\norm{\X\widehat{\bdelta}_k}_2^2
	\leq 6\rho_l^{-1}\lambda_k\sqrt{s_k}\norm{\widehat{\bdelta}_{J_k}}_2,
\end{align*}
where $\widehat{\bdelta}^{(1)}_{J_{k}^c}$ is a
subvector of $\widehat{\bdelta}_{J_{k}^c}$ consisting of the $s_k$ largest components in magnitude.  This leads to
\begin{align}%
	\norm{\widehat{\bdelta}_{J_k}}_2
	 & \leq 6\rho_l^{-1}\lambda_k\sqrt{s_k}, \label{im:th:ineq8-1} \\
	\norm{\widehat{\bdelta}^{(1)}_{J_{k}^c}}_2
	 & \leq 6\rho_l^{-1}\lambda_k\sqrt{s_k}. \label{im:th:ineq8-2}
\end{align}
Recall that $\widehat{\bdelta}_k=\widehat{\u}_k - \u_k^*$
and $\lambda_k = 4c\gamma_{d}^{-1}B_n$. In view of \eqref{im:th:ineq6} and \eqref{im:th:ineq8-1}, we have
\begin{align}\label{eq: A23-Li}
	     & n^{-1/2}\norm{\X(\widehat{\u}_k - \u_k^*)}_2
	=n^{-1/2}\norm{\X\widehat{\bdelta}_k}_2
	\leq  \left(6\lambda_k\sqrt{s_k}\norm{\widehat{\bdelta}_{J_k}}_2\right)^{1/2}\nonumber \\
	\leq & 6\rho_l^{-1/2}\lambda_k\sqrt{s_k}
	%=24c\rho_l^{-1/2}\gamma_{d}^{-1}B_n\sqrt{s_k}
	=24c\frac{\sqrt{s_k}}{\gamma_{d}\sqrt{\rho_l}}B_n
	\lesssim \frac{\sqrt{s_k}}{\gamma_{d}\sqrt{\rho_l}}B_n.
\end{align}

{\bf \underline{An upper bound for $\norm{\widehat{\u}_k - \u_k^*}_2$}}.
Let $\widehat{\bdelta}^{(2)}_{J_{k}^c}$ is a
subvector of $\widehat{\bdelta}_{J_{k}^c}$ excluding those components with the $s_k$ largest magnitude. Since the $j$th largest absolute component of $\widehat{\bdelta}_{J_{k}^c}$ is bounded from above by $\norm{\widehat{\bdelta}_{J_{k}^c}}_1/j$, we have
\begin{align*}
	\norm{\widehat{\bdelta}^{(2)}_{J_{k}^c}}_2^2 \leq \sum_{j=s_k+1}^{p}\norm{\widehat{\bdelta}_{J_k^c}}_1^2/j^2
	\leq s_k^{-1}\norm{\widehat{\bdelta}_{J_k^c}}_1^2.
\end{align*}
This inequality, together with \eqref{im:th:ineq5} and the Cauchy–Schwartz inequality, entails that
$\norm{\widehat{\bdelta}^{(2)}_{J_{k}^c}}_2 \leq s_k^{-1/2}\norm{\widehat{\bdelta}_{J_k^c}}_1
	\leq 3s_k^{-1/2}\norm{\widehat{\bdelta}_{J_k}}_1 \leq 3\norm{\widehat{\bdelta}_{J_k}}_2$.
Combining this with \eqref{im:th:ineq8-1} and \eqref{im:th:ineq8-2}, we have
\begin{align*}
	\norm{\widehat{\bdelta}_k}_2
	\leq \norm{\widehat{\bdelta}_{J_k}}_2 + \norm{\widehat{\bdelta}^{(1)}_{J_{k}^c}}_2
	+ \norm{\widehat{\bdelta}^{(2)}_{J_{k}^c}}_2
	\leq 4\norm{\widehat{\bdelta}_{J_k}}_2 + \norm{\widehat{\bdelta}^{(1)}_{J_{k}^c}}_2
	\leq 30\rho_l^{-1}\lambda_k\sqrt{s_k},
\end{align*}
which entails that
\begin{align}\label{eq: A24-Li}
	\norm{\widehat{\u}_k - \u_k^*}_2
	\leq 120c\frac{\sqrt{s_k}}{\rho_l\gamma_{d}}B_n
	\lesssim \frac{\sqrt{s_k}}{\rho_l\gamma_{d}}B_n,
\end{align}
since $\widehat{\bdelta}_k=\widehat{\u}_k - \u_k^*$
and $\lambda_k = 4c\gamma_{d}^{-1}B_n$.

	{\bf \underline{An upper bound for $\abs{\widehat{d}_k - d_k^*}/\sqrt{q}$}}.  Note that $\widehat{d}_k=\widetilde{d}_{k}/\sqrt{n}$. By \eqref{eq: A14-Li} from Lemma \ref{incomp:lem2}, conditional on the event $\Omega$, we have
\begin{align}\label{eq: A25-Li}
	\abs{\widehat{d}_k - d_k^*}/\sqrt{q}
	\leq \abs{\widetilde{d}_k - \sqrt{n}d_k^*}/\sqrt{nq}
	\lesssim B_n.
\end{align}

{\bf \underline{An upper bound for $\norm{\X(\widehat{\C}_{k}-\C^*_{k})}_{F}/\sqrt{nq}$}}.
Recall that $\widehat{\C}_{k}=\widehat{d}_{k}\widehat{\u}_{k}\widehat{\v}_{k}^T$ and $\C^*_{k}=d_{k}^*\u_k^*\v_k\strans$. We can write
\begin{align*}
	   \X(\widehat{\C}_{k}-\C^*_{k})
	= \widehat{d}_{k}\X\widehat{\u}_{k}\widehat{\v}_{k}\trans - d_{k}^*\X\u_k^*\v_k\strans %\\
	=  \bT_{1} + d_{k}^*(\bT_{2} + \bT_{3}  + \bT_{4})
	+ (\widehat{d}_k - d_k^*)(\bT_{2} + \bT_{3}  + \bT_{4}),
\end{align*}
where
\begin{align*}
	 & \bT_{1} = (\widehat{d}_{k} - d_{k}^*)\X\u_{k}^*\v_{k}\strans,~ \bT_{2}
	= \X(\widehat{\u}_{k}-\u_{k}^*)\v_{k}\strans,                             \\
	 & \bT_{3} = \X\u_{k}^*(\widehat{\v}_{k}-\v_{k}^*)\trans,
	~ \bT_{4} =  \X(\widehat{\u}_{k}-\u_{k}^*)(\widehat{\v}_{k}-\v_{k}^*)\trans.
\end{align*}
Thus, we have
\begin{align}\label{eq: A26-Li}
	      \frac{\norm{\X(\widehat{\C}_{k}-\C^*_{k})}_{F}}{\sqrt{nq}}%\nonumber \\
	\leq  \frac{\norm{\bT_{1}}_{F}}{\sqrt{nq}}
	+ \left(\frac{d_{k}^*}{\sqrt{q}}+ \frac{|\widehat{d}_k - d_k^*|}{\sqrt{q}}\right) \left(\frac{\norm{\bT_{2}}_{F}}{\sqrt{n}}
	+\frac{\norm{\bT_{3}}_{F}}{\sqrt{n}}
	+\frac{\norm{\bT_{4}}_{F}}{\sqrt{n}}\right).
\end{align}

We next find the bounds for $\norm{\bT_{1}}_{F}$, $\norm{\bT_{2}}_{F}$, $\norm{\bT_{3}}_{F}$, and $\norm{\bT_{4}}_{F}$ separately. Since $\norm{n^{-1/2}\X\u_{k}^*}_{2}=1$ and $\norm{\v_{k}^*}_{2}=1$, we have
\begin{align}\label{eq: T1-bound}
	\frac{\norm{\bT_{1}}_{F}}{\sqrt{nq}} = \frac{\abs{\widehat{d}_{k} - d_{k}^*}}{\sqrt{q}}\norm{n^{-1/2}\X\u_{k}^*}_{2}\norm{\v_{k}^*}_{2}
	= \frac{\abs{\widehat{d}_{k} - d_{k}^*}}{\sqrt{q}}
	\lesssim B_{n},
\end{align}
where the last inequality follows from \eqref{eq: A25-Li}. Similarly, using
$\norm{n^{-1/2}\X\u_{k}^*}_{2}=1$, $\norm{\v_{k}^*}_{2}=1$, \eqref{eq: A23-Li}
and \eqref{eq: A13-Li}, we obtain
\begin{align}
	\frac{\norm{\bT_{2}}_{F}}{\sqrt{n}}
	 & = n^{-1/2}\norm{\X(\widehat{\u}_{k}-\u_{k}^*)}_{2}\norm{\v_{k}^*}_{2}
	\lesssim\frac{\sqrt{s_k}}{\gamma_{d}\sqrt{\rho_l}}B_n,\label{eq: T2-bound}                  \\
	\frac{\norm{\bT_{3}}_{F}}{\sqrt{n}}
	 & = n^{-1/2}\norm{\X\u_{k}^*}_{2}\norm{\widetilde{\v}_{k}-\v_{k}^*}_{2}
	\lesssim\gamma_{d}^{-1}B_n,\label{eq: T3-bound}                                             \\
	\frac{\norm{\bT_{4}}_{F}}{\sqrt{n}}
	 & = n^{-1/2}\norm{\X(\widehat{\u}_{k}-\u_{k}^*)}_{2}\norm{\widetilde{\v}_{k}-\v_{k}^*}_{2}
	\lesssim\frac{\sqrt{s_k}}{\gamma_{d}^2\sqrt{\rho_l}}B_n^2. \label{eq: T4-bound}
\end{align}
If follows from Condition \ref{cond2} and \eqref{eq: A25-Li} that
\begin{align}\label{eq: dk-hat-bound}
	\frac{d_{k}^*}{\sqrt{q}}+ \frac{|\widehat{d}_k - d_k^*|}{\sqrt{q}}
	\leq c_3 + c_4B_n,
\end{align}
where $c_3$ and $c_4$ are some positive constants.  Combining this with \eqref{eq: Bn-bound} and \eqref{eq: A26-Li}-\eqref{eq: T4-bound} entails
\begin{align}\label{eq: A31-Li}
	          \frac{1}{\sqrt{nq}}\norm{\X(\widehat{\C}_{k}-\C^*_{k})}_{F} %\nonumber\\                                                                                                                                   
	\lesssim & B_{n} + (c_3 + c_4B_n)\left(\frac{\sqrt{s_k}}{\gamma_{d}\sqrt{\rho_{l}}}B_{n} + \frac{B_{n}}{\gamma_{d}} + \frac{\sqrt{s_k}}{\gamma_{d}^{2}\sqrt{\rho_{l}}}B_{n}^{2}\right) \nonumber                   \\
	=        & \left(1+\frac{c_3\sqrt{s_k}}{\gamma_{d}\sqrt{\rho_{l}}} + \frac{c_3}{\gamma_{d}}\right)B_{n} +
	\left(\frac{c_3\sqrt{s_k}}{\gamma_{d}^2\sqrt{\rho_{l}}}+\frac{c_4\sqrt{s_k}}{\gamma_{d}\sqrt{\rho_{l}}} + \frac{c_4}{\gamma_{d}}\right)B_{n}^2+ \frac{c_4\sqrt{s_k}}{\gamma_{d}^{2}\sqrt{\rho_{l}}}B_n^3 \nonumber \\
	\leq     & \frac{c_5\sqrt{s_{k}}}{\gamma_{d}\sqrt{\rho_{l}}}B_n + \frac{c_6\sqrt{s_{k}}}{\gamma_{d}\sqrt{\rho_{l}}}B_n^2+ \frac{c_4\sqrt{s_k}}{\gamma_{d}^{2}\sqrt{\rho_{l}}}B_n^3                                 %\nonumber\\
	\lesssim \frac{\sqrt{s_{k}}}{\gamma_{d}\sqrt{\rho_{l}}}B_n ,
\end{align}
where $c_5$ and $c_6$ are some positive constants.

	{\bf \underline{An upper bound for $\norm{\widehat{\C}_{k}-\C^*_{k}}_{F}/\sqrt{q}$}}.
Similar to \eqref{eq: A26-Li}, we can show that
\begin{align*}%\label{eq: Ck-error}
	\frac{\norm{\widehat{\C}_{k}-\C^*_{k}}_{F}}{\sqrt{q}} %\nonumber\\
	\leq  \frac{\norm{\bT_{5}}_{F}}{\sqrt{q}}
	+ \left(\frac{d_{k}^*}{\sqrt{q}}+ \frac{|\widehat{d}_k - d_k^*|}{\sqrt{q}}\right) \left(\norm{\bT_{6}}_{F}+\norm{\bT_{7}}_{F}+\norm{\bT_{8}}_{F}\right),
\end{align*}
where
\begin{align*}
	 & \bT_{5} = (\widehat{d}_{k} - d_{k}^*)\u_{k}^*\v_{k}\strans,~ \bT_{6}
	= (\widehat{\u}_{k}-\u_{k}^*)\v_{k}\strans,                             \\
	 & \bT_{7} = \u_{k}^*(\widehat{\v}_{k}-\v_{k}^*)\trans,
	~ \bT_{8} =  (\widehat{\u}_{k}-\u_{k}^*)(\widehat{\v}_{k}-\v_{k}^*)\trans.
\end{align*}
Recall that $J_k$ is the support of
$\u_k^*$, and $\u^*_{J_k}$ and $\u^*_{J_k^c}$ are the subvectors of $\u^*_k$ formed by components in $J_k$ and $J_k^c$, respectively. Then we have
$0=\norm{\u^*_{J_k^c}}_1\leq 3\norm{\u^*_{J_k}}_1$. It follows from $s_k=\norm{\u_k^*}_0\leq s$ and Condition  \ref{cond4} that $\norm{\u^*_{J_k}}_{2}
	\leq \rho_l^{-1/2} n^{-1/2}\norm{\X\u_k^*}_{2}$.
Since $\norm{n^{-1/2}\X\u_{k}^*}_{2}=1$ and $\norm{\u^*_{k}}_{2}=\norm{\u^*_{J_k}}_{2}$, we have
\begin{align*}
	\norm{\u^*_{k}}_{2}=\norm{\u^*_{J_k}}_{2}
	\leq \rho_l^{-1/2} n^{-1/2}\norm{\X\u_k^*}_{2} = \rho_l^{-1/2}.
\end{align*}
Using similar arguments for bounding
$\norm{\X(\widehat{\C}_{k}-\C^*_{k})}_{F}/\sqrt{nq}$, we can obtain a similar bound
\begin{align}\label{eq: A32-Li}
	\frac{1}{\sqrt{q}}\norm{\widehat{\C}_{k}-\C^*_{k}}_{F} \lesssim \frac{\sqrt{s_k}}{\rho_l\gamma_{d}}B_n.
\end{align}
It concludes the proof of Theorem \ref{im:th}.

\subsection{Proof of Proposition \ref{com:th}}\label{App-com-th}

Recall that the event $\mathcal{B}$, defined in \eqref{eq: eventB} in the proof of Lemma \ref{comp:lem1}, holds with probability at least $1-2e^{-(\sqrt{n}+\sqrt{q})^2}$,
that is, $P(\mathcal{B})\geq 1-2e^{-(\sqrt{n}+\sqrt{q})^2}$.
Thus, to prove Proposition \ref{com:th}, it suffices to show that, conditional on the event $\mathcal{B}$, the following six inequalities
\begin{align*}
	n^{-1/2}\norm{\X(\widehat{\u}_k - \u_k^*)}_2
	 & \lesssim \frac{\sqrt{s_k}}{\gamma_{d}\sqrt{\rho_l}}\widetilde{B}_n, \quad\,
	\norm{\widehat{\u}_k - \u_k^*}_2  \lesssim \frac{\sqrt{s_k}}{\rho_l\gamma_{d}}\widetilde{B}_n, \\
	\norm{\widehat{\v}_k - \v_k^*}_2
	 & \lesssim \gamma_{d}^{-1}\widetilde{B}_n,\quad\,
	\abs{\widehat{d}_k - d_k^*}/\sqrt{q} \lesssim \widetilde{B}_n,                                 \\
	\frac{1}{\sqrt{nq}}\norm{\X(\widehat{\C}_{k}-\C^*_{k})}_{F}
	 & \lesssim \frac{\sqrt{s_k}}{\gamma_{d}\sqrt{\rho_l}}\widetilde{B}_n,\quad\,
	\frac{1}{\sqrt{q}}\norm{\widehat{\C}_{k}-\C^*_{k}}_{F} \lesssim \frac{\sqrt{s_k}}{\rho_l\gamma_{d}}\widetilde{B}_n
\end{align*}
hold uniformly over $k=1, \cdots, r^*$.

Hereafter our analysis will be conditional on the event $\mathcal{B}$.  By \eqref{eq: B4-Li}, %from Lemma \ref{comp:lem1}, 
we have that, conditional on the event $\mathcal{B}$, $	\norm{\widehat{\v}_k - \v_k^*}_2 \lesssim \gamma_{d}^{-1}\widetilde{B}_n$ holds uniformly over $k=1, \cdots, r^*$.
Using similar arguments for proving
\eqref{eq: A23-Li}, \eqref{eq: A24-Li}, \eqref{eq: A25-Li}, \eqref{eq: A31-Li}, and \eqref{eq: A32-Li}, we can show that conditional on the event $\mathcal{B}$, other five inequalities also holds uniformly over $k=1, \cdots, r^*$. So the details are omitted here to save space.
This completes the proof of Proposition \ref{com:th}.

\subsection{Proof of Theorem \ref{th3}}\label{App-th3}

%{\bf Proof}. 
Under Conditions \ref{cond1}, \ref{cond3}, and \ref{cond5}, the event $\Omega =\left\{\norm{\widetilde{\bDelta}}_F/\sqrt{nq}\leq cB_n\right\}$ following from Lemma \ref{incomp:lem1}
holds with probability at least $1-(n+q)^{-1}$, where $c$ is a positive constant. Thus it suffices to show that $\widehat{r}=r^*$ conditional on $\Omega$.

Without loss of generality, $B_n$ can be simplified as $O\left(\sqrt{m^{-1}r(n \vee q)\log(n+q)}\right)$ when assume $n\geq 2$. Since $\widetilde{d}_k/\sqrt{nq}$ and $d_k^*/\sqrt{q}$ are the singular values of $\widetilde{\Y}/\sqrt{nq}$ and $\X\C^*/\sqrt{nq}$,
respectively, applying Weyl's theorem \citep[Theorem 2]{stewart1998} leads to
\begin{align}\label{eq: B2}
	\abs{\widetilde{d}_k - \sqrt{n}d_k^*}/\sqrt{nq}
	=\abs{\widetilde{d}_k/\sqrt{nq} - d_k^*/\sqrt{q}}
	\leq \norm{\widetilde{\bDelta}}_F/\sqrt{nq}\,\,\,\,\text{for}\,\,1\leq k\leq r^{*}
\end{align}
%for $1\leq k\leq r^{*}$ 
and
\begin{align}\label{eq: B3}
	\abs{\widetilde{d}_k}/\sqrt{nq}
	=\abs{\widetilde{d}_k/\sqrt{nq} - 0}
	\leq \norm{\widetilde{\bDelta}}_F/\sqrt{nq}\,\,\,\,\text{for}\,\,k>r^{*}.
\end{align}
%for $k>r^{*}$.

From now on, we condition on the event $\Omega$. Then we have
\begin{align}\label{eq: B4}
	\frac{\norm{\widetilde{\bDelta}}_F}{\sqrt{nq}}
	\leq cB_n
	=O\left(\sqrt{m^{-1}r(n \vee q)\log(n+q)}\right)
	=O\left(\frac{1}{\log n}\right),
\end{align}
where the last identity follows from the assumption that $m^{-1}r(n\vee q)(\log^2n)\log (n+q)=O(1)$.
We can show that conditional on the event $\Omega$, for sufficiently largely $n$, we have that $(nq)^{-1/2}(\widetilde{d}_k-\widetilde{d}_{k+1})$ is greater than $\tau_n$ for all $1\leq k\leq r^{*}$ and smaller than $\tau_n$ for $k> r^{*}$.
To this end, we consider three cases.

Case 1: $1\leq k\leq  r^{*}-1$. Note that
\begin{align*}
	   (nq)^{-1/2}(\widetilde{d}_k-\widetilde{d}_{k+1})        
	=  q^{-1/2}(d^{*}_k-d^{*}_{k+1})+(nq)^{-1/2}(\widetilde{d}_k-\sqrt{n}d^{*}_k) + (nq)^{-1/2}(\sqrt{n}d^{*}_{k+1}-\widetilde{d}_{k+1}).
\end{align*}
Thus, it follows from Condition \ref{cond2}, \eqref{eq: B2} and \eqref{eq: B4} that
\begin{align*}%\label{eq: case1-bound}
	(nq)^{-1/2}(\widetilde{d}_k-\widetilde{d}_{k+1})
	\geq  q^{-1/2}(d^{*}_k-d^{*}_{k+1})-2\norm{\widetilde{\bDelta}}_F/\sqrt{nq}%\nonumber \\
	%\geq&  \gamma_d-O\left(\sqrt{m^{-1}r(n \vee q)\log(n+q)}\right)
	\geq  \gamma_d-O\left(\frac{1}{\log n)}\right)
	> \frac{\log\log n}{\log n}%=\tau_n
\end{align*}
for sufficiently largely $n$.

Case 2: $k=r^{*}$. In view of Condition \ref{cond2}, \eqref{eq: B3} and \eqref{eq: B4}, we have
\begin{align*}%\label{eq: case2-bound}
	     & (nq)^{-1/2}(\widetilde{d}_k-\widetilde{d}_{k+1})
	=q^{-1/2}d^{*}_k+(nq)^{-1/2}(\widetilde{d}_k-\sqrt{n}d^{*}_k)-(nq)^{-1/2}\widetilde{d}_{k+1}\nonumber \\
	\geq & O(1)-2\norm{\widetilde{\bDelta}}_F/\sqrt{nq}                                                   %\\
	%\geq&  \gamma_d-O\left(\sqrt{m^{-1}r(n \vee q)\log(n+q)}\right)
	\geq O(1)-O\left(\frac{1}{\log n}\right)
	> \frac{\log\log n}{\log n}%=\tau_n
\end{align*}
for sufficiently largely $n$.

Case 3: $k> r^{*}$. It follows from \eqref{eq: B3} and \eqref{eq: B4} that
\begin{align*}%\label{eq: case3-bound}
	(nq)^{-1/2}(\widetilde{d}_k-\widetilde{d}_{k+1})
	\leq  \abs{\widetilde{d}_k}/\sqrt{nq} + \abs{\widetilde{d}_{k+1}}/\sqrt{nq}
	\leq 2\norm{\widetilde{\bDelta}}_F/\sqrt{nq}%\nonumber                       %\\
	\leq  O\left(\frac{1}{\log n}\right)
	< \frac{\log\log n}{\log n}%=\tau_n
\end{align*}
for sufficiently largely $n$.

Combining Cases 1-3 above along with $\tau_n=(\log n)^{-1}\log\log n$ yields that, conditional on the event $\Omega$, the following bounds hold for sufficiently large $n$:
\begin{align*}
	 & (nq)^{-1/2}(\widetilde{d}_k-\widetilde{d}_{k+1})
	>\tau_n\,\,\text{for}\,\,1\leq k\leq r^{*};         \\
	 & (nq)^{-1/2}(\widetilde{d}_k-\widetilde{d}_{k+1})
	<\tau_n\,\,\text{for}\,\,k>r^{*}.
\end{align*}
Therefore, by choosing $\widehat{r}=\arg\max_{k}\{1\leq k\leq r: (nq)^{-1/2}(\widetilde{d}_k-\widetilde{d}_{k+1})>\tau_n\}$, we have $r=r^{*}$ with probability at least $1-(n+q)^{-1}$ for sufficiently large $n$, which concludes the proof of Theorem \ref{th3}.

\section{Proofs of Lemmas}\label{App-B}

\subsection{Proof of Lemma \ref{incomp:lem1}}\label{App-incomp-lem1}

Recall that $\widetilde{\Y}$ is the minimizer of \eqref{opt:matrix-completion}. Thus, we have
\begin{equation}\label{eq:A1-add}
	m^{-1}\norm{\proj_\M(\Y) - \proj_\M(\widetilde{\Y})}_F^2 \leq
	m^{-1}\norm{\proj_\M(\Y) - \proj_\M(\X\C^*)}_F^2.
\end{equation}
Note that $\proj_\M(\Y) - \proj_\M(\X\C^*) = \proj_\M(\E)$ and
\begin{align*}
	\proj_\M(\Y) - \proj_\M(\widetilde{\Y})
	=  [\proj_\M(\Y) - \proj_\M(\X\C^*)]
	-[\proj_\M(\widetilde{\Y}) - \proj_\M(\X\C^*)]%\nonumber \\
	=  \proj_\M(\E)
	- \proj_\M(\widetilde{\bDelta}),
\end{align*}
where $\widetilde{\bDelta} = \widetilde{\Y} - \X\C^*$.
Substituting these two identities into \eqref{eq:A1-add} yields
\begin{equation*}%\label{incomp:lem1:ineq1}
	m^{-1}\norm{\proj_\M(\widetilde{\bDelta})}_F^2 \leq 2m^{-1}\inner{\proj_\M(\E), \proj_\M(\widetilde{\bDelta})}.
\end{equation*}
For simplicity, we write
$\widetilde{\bDelta}_{\M}=\proj_\M(\widetilde{\bDelta})$ and
$\E_\M=\proj_\M(\E)$.  Thus the above inequality can be written as
\begin{align}\label{incomp:lem1:ineq1}
	m^{-1}\norm{\widetilde{\bDelta}_{\M}}_F^2 \leq 2m^{-1}\inner{\E_\M, \widetilde{\bDelta}_{\M}}.
\end{align}
Let $\widetilde{\bDelta}_{ij}$ be the $(i, j)$ entry of the matrix $\widetilde{\bDelta}$. By the definition of of $\widetilde{\Y}$, we have $\norm{\widetilde{\Y}}_{\max}\leq L$. Under the assumption that $\norm{\X\C^*}_{\max} \leq L$, we further have
\begin{align*}
	\norm{\widetilde{\bDelta}}_{\max}
	=\norm{\widetilde{\Y} - \X\C^*}_{\max}
	\leq \norm{\widetilde{\Y}}_{\max} + \norm{\X\C^*}_{\max} \leq 2L.
\end{align*}

Recall that $\M=\{(i, j): y_{ij}\,\,\text{is observed},\,\, 1\leq i\leq n, 1\leq j\leq q\}$.
Denote by $\{\omega_t\}_{t=1}^m$ the sampled sequence of entries, where $\omega_t=(i_t,j_t)\in\M$ for all $t$. Then we can define a sequence of matrices $\{\W_t\}_{t=1}^m$, where the entries of $\W_t\in\R^{n\times q}$ are all zeros except for $1$ at the location $\omega_t$ (i.e. $(\W_t)_{i_t,j_t}=1$).
Let $\{\epsilon_t\}_{t=1}^{m}$ be a Rademacher sequence independent of
$(\omega_t, \W_t)_{t=1}^m$. Define $\bSigma_R = m^{-1}\sum_{t=1}^m\epsilon_t\W_t$.

In order to proceed, we first show that
\begin{align}\label{incomp:lem1:ineq5}
	\frac{1}{nq}\norm{\widetilde{\bDelta}}_F^2
	\leq c\max\left\{\mu L^2\sqrt{\frac{\log(n+q)}{m}}, \frac{rnq}{m^2}\mu^2d_1^2(\E_\M)+rnq\mu^2L^2\mathbb{E}^2\left[d_1(\bSigma_R)\right]\right\},
\end{align}
where $c$ is some positive constant.
%, and then bound both $d_1^2(\E_\M)$ and $\mathbb{E}\left[d_1(\bSigma_R)\right]$. To prove \eqref{incomp:lem1:ineq5}, we consider two cases.
%
To prove this, we consider two cases.

Case 1: $\sum_{i,j}\pi_{ij}\widetilde{\bDelta}_{ij}^2 < 4L^2\sqrt{\frac{64\log(n+q)}{m\log(6/5)}}$. Then by Condition \ref{cond3}, we have
\begin{align}\label{incomp:lem1:ineq2}
	\frac{1}{\mu nq}\norm{\widetilde{\bDelta}}_F^2 \leq \sum_{i,j}\pi_{ij}\widetilde{\bDelta}_{ij}^2
	< 4L^2\sqrt{\frac{64\log(n+q)}{m\log(6/5)}}.
\end{align}

Case 2: $\sum_{i,j}\pi_{ij}\widetilde{\bDelta}_{ij}^2 \geq 4L^2\sqrt{\frac{64\log(n+q)}{m\log(6/5)}}$. Note that
$\rank{\widetilde{\bDelta}} \leq \rank{\widetilde{\Y}} + \rank{\X\C^*} \leq 2r$. Applying Lemma 12 of \citet{klopp2014}
entails
\begin{align*}
	\frac{1}{m}(2L)^{-2}\norm{\widetilde{\bDelta}_\M}_F^2
	 & \geq \frac{1}{2}(2L)^{-2}\sum_{i,j}\pi_{ij}\widetilde{\bDelta}_{ij}^2 - 88\mu rnq
	\mathbb{E}^2\left[d_1(\bSigma_R)\right]                                              \\
	 & \geq \frac{1}{2\mu nq}(2L)^{-2}\norm{\widetilde{\bDelta}}_F^2 - 88\mu rnq
	\mathbb{E}^2\left[d_1(\bSigma_R)\right],
\end{align*}
where the last inequality follows from Condition \ref{cond3}.
Multiplying both sides of the above inequality by $4L^2$
yields
\begin{align}\label{incomp:lem1:ineq3}
	\frac{1}{m}\norm{\widetilde{\bDelta}_\M}_F^2 \geq \frac{1}{2\mu nq}\norm{\widetilde{\bDelta}}_F^2 - 352L^2\mu rnq
	\mathbb{E}^2\left[d_1(\bSigma_R)\right].
\end{align}
This inequality, together with \eqref{incomp:lem1:ineq1}, gives
\begin{align*}
	\frac{1}{2\mu nq}\norm{\widetilde{\bDelta}}_F^2
	 & \leq 2m^{-1}\inner{\E_\M, \widetilde{\bDelta}_\M}+352L^2\mu rnq\mathbb{E}^2\left[d_1(\bSigma_R)\right] \\
	 & = 2m^{-1}\inner{\E_\M, \widetilde{\bDelta}}+352L^2\mu rnq\mathbb{E}^2\left[d_1(\bSigma_R)\right].
\end{align*}
After some algebra, we obtain
\begin{align*}
	\frac{1}{2\mu nq}\norm{\widetilde{\bDelta}}_F^2
	 & \leq 2m^{-1}d_1(\E_\M)\norm{\widetilde{\bDelta}}_* + 352L^2\mu rnq\mathbb{E}^2\left[d_1(\bSigma_R)\right]           \\
	 & \leq 2m^{-1}d_1(\E_\M)\sqrt{2r}\norm{\widetilde{\bDelta}}_F + 352L^2\mu rnq\mathbb{E}^2\left[d_1(\bSigma_R)\right],
\end{align*}
where $\norm{\widetilde{\bDelta}}_*$ is nuclear norm of $\widetilde{\bDelta}$ (i.e., the sum of the singular values of $\widetilde{\bDelta}$).
Using the fact that $2xy \leq a^{-1}x^2 + ay^2$ for any $a > 0$, we have
\begin{equation*}
	\frac{1}{2\mu nq}\norm{\widetilde{\bDelta}}_F^2 \leq a^{-1}m^{-1}d_1^2(\E_\M) + 2m^{-1}ar\norm{\widetilde{\bDelta}}_F^2 +
	352L^2\mu rnq\mathbb{E}^2\left[d_1(\bSigma_R)\right].
\end{equation*}
Taking $a=m/(8r\mu nq)$ in the inequality above leads to
\begin{equation}\label{incomp:lem1:ineq4}
	\frac{1}{4\mu nq}\norm{\widetilde{\bDelta}}_F^2 \leq \frac{8r\mu nq}{m^2}d_1^2(\E_\M) + 352L^2\mu rnq\mathbb{E}^2\left[d_1(\bSigma_R)\right].
\end{equation}
Thus \eqref{incomp:lem1:ineq5} holds by combining \eqref{incomp:lem1:ineq2} in Case 1 and \eqref{incomp:lem1:ineq4} in Case 2.

Next, we derive the bounds for $d_1^2(\E_\M)$ and $\mathbb{E}\left[d_1(\bSigma_R)\right]$, respectively.
Under Conditions \ref{cond1} and \ref{cond3}, it follows from
Lemma 5 of \citet{klopp2014} that the event
\begin{align*}
	\mathcal{A}
	= \left\{\frac{1}{\sigma m}d_1(\E_\M) \leq c\max\left[\sqrt{\frac{\nu(t+\log(n+q))}{m(n\wedge q)}}, \frac{\log(n\wedge q)(t+\log(n+q))}{m}\right]\right\}
\end{align*}
holds with probability at least $1-e^{-t}$ for all $t>0$.
Taking $t=\log(n+q)$ gives that
%\begin{align}\label{incomp:lem1:ineq6}
%	\frac{1}{m}d_1(\E_\M)
%	 & \leq
%	c\max\left\{\sigma\sqrt{\frac{\nu\log(n+q)}{m(n\wedge q)}},~
%	\sigma\frac{\log(n\wedge q)\log(n+q)}{m}\right\}\nonumber \\
%	 & \leq c\sigma\sqrt{\frac{\nu\log(n+q)}{m(n\wedge q)}}
%\end{align}
\begin{align}\label{incomp:lem1:ineq6}
\frac{1}{m}d_1(\E_\M)
 \leq
c\max\left\{\sigma\sqrt{\frac{\nu\log(n+q)}{m(n\wedge q)}},~
\sigma\frac{\log(n\wedge q)\log(n+q)}{m}\right\}%\nonumber \\
\leq c\sigma\sqrt{\frac{\nu\log(n+q)}{m(n\wedge q)}}
\end{align}
holds with probability at least $1-(n+q)^{-1}$, where the second inequality follows from the assumption that $m\geq \nu^{-1}(n\wedge q)\log^{3}(n+q)$.

Applying Lemma 6 of \cite{klopp2014} with $m\geq \nu^{-1}(n\wedge q)\log^{3}(n+q)$ yields
\begin{align}\label{incomp:lem1:ineq7}
	\mathbb{E}\left[d_1(\bSigma_R)\right] \leq c\sqrt{\frac{\nu\log(n+q)}{m(n\wedge q)}}.
\end{align}
In view of \eqref{incomp:lem1:ineq5}, \eqref{incomp:lem1:ineq6}
and \eqref{incomp:lem1:ineq7}, we have
\begin{align*}%\label{incomp:lem1:ineq8}
	 & \frac{1}{nq}\norm{\widetilde{\bDelta}}_F^2
	\leq c\max\left\{\mu L^2\sqrt{\frac{\log(n+q)}{m}},~
	rnq\mu^2(\sigma^2+L^2)\frac{\nu\log(n+q)}{m(n\wedge q)}\right\}\nonumber                                                \\
	 & \leq c\max\left\{\mu L^2\sqrt{\frac{\log(n+q)}{m}},~\nu\mu^2(\sigma^2\vee L^2)\frac{r(n \vee q)\log(n+q)}{m}\right\}
	=cB_n^2
\end{align*}
holds with probability at least $1-(n+q)^{-1}$, where second inequality follows from the fact that $nq=(n\wedge q)(n \vee q)$ and
$\sigma^2+L^2\leq 2(\sigma^2\vee L^2)$.	This completes the proof of Lemma \ref{incomp:lem1}.

\subsection{Proof of Lemma \ref{incomp:lem2}}\label{App-incomp-lem2}

Recall that $\widetilde{\z}_k$ and $\widetilde{\v}_k$ are the $k$th left and right singular vectors of $\widetilde{\Y}/\sqrt{nq}$, respectively.
In addition, $n^{-1/2}\X\u_k^*$ and $\v_k^*$ are the $k$th left and right singular vectors of $\X\C^*/\sqrt{nq}$.
Using the fact that $\norm{\widetilde{\bDelta}}_{op}\leq \norm{\widetilde{\bDelta}}_F$ and Theorem 3 of \citet{yu2014} yields
\begin{align}\label{eq: incomp-add1-1}
	  & \norm{\widetilde{\z}_k - n^{-1/2}\X\u_k^*}_2
	\leq \frac{2^{3/2}(2d_1^*/\sqrt{q}+\norm{\widetilde{\bDelta}}_F/\sqrt{nq})\norm{\widetilde{\bDelta}}_F/\sqrt{nq}}
	{\min(d_{k-1}^{*2} - d_k^{*2}, d_{k}^{*2} - d_{k+1}^{*2})/q} \nonumber         \\
	= & \frac{2^{3/2}}{\min(d_{k-1}^{*2} - d_k^{*2}, d_{k}^{*2} - d_{k+1}^{*2})/q}
	\left(\frac{2d_1^*}{\sqrt{q}}\cdot\frac{\norm{\widetilde{\bDelta}}_F}{\sqrt{nq}}
	+\frac{\norm{\widetilde{\bDelta}}_F^2}{nq}\right)
	% \lesssim \gamma_{d}^{-1}\norm{\widetilde{\bDelta}}_F/\sqrt{nq}
\end{align}
for each $k=1, \cdots, r^*$, where $d_{0}^{*2}=+\infty$, $d_{r^*+1}^{*2}=-\infty$, and $\widetilde{\bDelta}=\widetilde{\Y}-\X\C^*$.
This together with Condition \ref{cond2} entails
\begin{align}\label{eq: incomp-add1}
	\norm{\widetilde{\z}_k - n^{-1/2}\X\u_k^*}_2 \lesssim \gamma_d^{-1}\left[\norm{\widetilde{\bDelta}}_F/\sqrt{nq} + \norm{\widetilde{\bDelta}}_F^2/(nq)\right].
\end{align}
Similarly to above, we have
\begin{align}\label{eq: incomp-add2}
	\norm{\widetilde{\v}_k - \v_k^*}_2 \lesssim \gamma_d^{-1}\left[\norm{\widetilde{\bDelta}}_F/\sqrt{nq} + \norm{\widetilde{\bDelta}}_F^2/(nq)\right].
\end{align}

Note that $\widetilde{d}_k/\sqrt{nq}$ and $d_k^*/\sqrt{q}$ are the singular values of $\widetilde{\Y}/\sqrt{nq}$ and $\X\C^*/\sqrt{nq}$,
respectively. Thus, an application of Weyl's theorem \citep[Theorem 2]{stewart1998} leads to
\begin{align}\label{eq: incomp-add3}
	\abs{\widetilde{d}_k - \sqrt{n}d_k^*}/\sqrt{nq}
	=\abs{\widetilde{d}_k/\sqrt{nq} - d_k^*/\sqrt{q}}
	\leq \norm{\widetilde{\bDelta}}_F/\sqrt{nq}.
\end{align}
Combining \eqref{eq: incomp-add1}, \eqref{eq: incomp-add2}, and \eqref{eq: incomp-add3} together yields that the following three inequalities
\begin{align}
	\norm{\widetilde{\z}_k - n^{-1/2}\X\u_k^*}_2
	 & \lesssim \gamma_d^{-1}\left[\norm{\widetilde{\bDelta}}_F/\sqrt{nq} + \norm{\widetilde{\bDelta}}_F^2/(nq)\right], \label{eq: A12-Li-old} \\
	\norm{\widetilde{\v}_k - \v_k^*}_2
	 & \lesssim \gamma_d^{-1}\left[\norm{\widetilde{\bDelta}}_F/\sqrt{nq} + \norm{\widetilde{\bDelta}}_F^2/(nq)\right], \label{eq: A13-Li-old} \\
	\abs{\widetilde{d}_k - \sqrt{n}d_k^*}/\sqrt{nq}
	 & \leq \norm{\widetilde{\bDelta}}_F/\sqrt{nq}, \label{eq: A14-Li-old}
\end{align}
hold uniformly for all $k=1, \cdots, r^*$.

Recall that the event $\Omega =\left\{\norm{\widetilde{\bDelta}}_F/\sqrt{nq}\leq cB_n\right\}$. It follows from Lemma \ref{incomp:lem1} that
$\norm{\widetilde{\bDelta}}_F/\sqrt{nq}
	\leq cB_n$
holds with probability at least $1-(n+q)^{-1}$, where $c$ is a positive constant. Then we have $P(\Omega)\geq 1-(n+q)^{-1}$.

Under the assumption that $m\geq \max\{\nu^{-1}(n\wedge q)\log^{3}(n+q), r(n\vee q)\log(n+q)\}$, we have
\begin{align*}
	\frac{\log(n+q)}{m}\leq\frac{\nu\log(n+q)}{(n\wedge q)\log^{3}(n+q)}\leq \frac{\nu}{\log^2(2)}\,\,\,
	\mbox{and}\,\,\, \sqrt{\frac{r(n \vee q)\log(n+q)}{m}}\leq 1.%,
\end{align*}
Thus, by the definition of $B_n$, combining these two bounds yields
\begin{align}\label{eq: Bn-bound}
	B_n \leq \max\left\{L\mu^{1/2}\nu^{1/4}/\log^{1/2}(2),\,\mu(\sigma\vee L)\nu^{1/2} \right\}
	\leq c,
\end{align}
where $c$ is some positive constant.
This, together with \eqref{eq: A12-Li-old}, \eqref{eq: A13-Li-old}, and \eqref{eq: A14-Li-old}, entails that, conditional on the event $\Omega$, the following inequalities
\begin{align}
	\norm{\widetilde{\z}_k - n^{-1/2}\X\u_k^*}_2
	 & \lesssim \gamma_d^{-1}(B_n+B_n^2)
	=\gamma_d^{-1}(1+B_n)B_n\lesssim \gamma_d^{-1}B_n, \label{eq: A12-Li} \\
	\norm{\widetilde{\v}_k - \v_k^*}_2
	 & \lesssim \gamma_d^{-1}(B_n+B_n^2)
	=\gamma_d^{-1}(1+B_n)B_n\lesssim \gamma_d^{-1}B_n, \label{eq: A13-Li} \\
	\abs{\widetilde{d}_k - \sqrt{n}d_k^*}/\sqrt{nq}
	 & \leq B_n, \label{eq: A14-Li}
\end{align}
hold uniformly for all $k=1, \cdots, r^*$.  Since $P(\Omega)\geq 1-(n+q)^{-1}$, we have that these three inequalities hold uniformly for all $k=1, \cdots, r^*$ with probability at least $1-(n+q)^{-1}$. This concludes the proof of Lemma \ref{incomp:lem2}.

\subsection{Proof of Lemma \ref{comp:lem1}}\label{App-comp-lem1}

Let $d_1(\E)$ be the the largest singular value of $\E$. Note that the operator norm of $\E$ is equal to its largest singular value, that is, $\norm{\E}_{op}=d_1(\E)$.
Recall that $\widetilde{d}_k/\sqrt{nq}$, $\widetilde{\z}_k$ and $\widetilde{\v}_k$ are the $k$th singular value, left and right singular vectors of $\Y/\sqrt{nq}$, respectively. In addition, $n^{-1/2}\X\u_k^*$ and $\v_k^*$ are the $k$th left and right singular vectors of $\X\C^*/\sqrt{nq}$.
Using Theorem 3 of \citet{yu2014} and Condition \ref{cond2} yields
\begin{align*}
	  & \norm{\widetilde{\z}_k - n^{-1/2}\X\u_k^*}_2 \leq
	\frac{2^{3/2}(2d_1^*/\sqrt{q} + \norm{\E}_{op}/\sqrt{nq})\norm{\E}_{op}/\sqrt{nq}}{\min(d_{k-1}^{*2} - d_k^{*2}, d_{k}^{*2} - d_{k+1}^{*2})/q} \nonumber \\
	= & \frac{2^{3/2}}{\min(d_{k-1}^{*2} - d_k^{*2}, d_{k}^{*2} - d_{k+1}^{*2})/q}
	\left(\frac{2d_1^*}{\sqrt{q}}\cdot\frac{d_1(\E)}{\sqrt{nq}}
	+\frac{d_1^2(\E)}{nq}\right)
\end{align*}
for each $k=1, \cdots, r^*$, where $d_{0}^{*2}=+\infty$, $d_{r^*+1}^{*2}=-\infty$ and we use $\Y-\X\C^*=\E$.
This together with Condition \ref{cond2} entails
\begin{align}\label{eq: B1-Li}
	\norm{\widetilde{\z}_k - n^{-1/2}\X\u_k^*}_2
	%	\lesssim \gamma_d^{-1}\left[\norm{\E}_{op}/\sqrt{nq} + \norm{\E}_{op}^2/(nq)\right] \\
	\lesssim \gamma_d^{-1}\left[d_1(\E)/\sqrt{nq} + d_1^2(\E)/(nq)\right].
\end{align}
Similarly, we have
\begin{align}
	 & \norm{\widetilde{\v}_k - \v_k^*}_2 \lesssim
	\gamma_d^{-1}\left[d_1(\E)/\sqrt{nq} + d_1^2(\E)/(nq)\right]\label{eq: B2-Li}
\end{align}
for each $k=1, \cdots, r^*$.

We next find an upper bound for $d_1(\E)$. It follows from Condition \ref{cond1} and Proposition 2.4 of \citet{rudelson2010non} that
\begin{equation*}
	P(\sigma^{-1}d_1(\E) > C_1(\sqrt{n}+\sqrt{q}) + t) \leq 2e^{-C_2t^2},
\end{equation*}
where $C_1$ and $C_2$ are some positive constants. Taking $t=(\sqrt{n}+\sqrt{q})/\sqrt{C_2}$, we have that the event
\begin{align}\label{eq: eventB}
	\mathcal{B}=\left\{d_1(\E) \leq  C_3\sigma(\sqrt{n}+\sqrt{q})\right\}
\end{align}
holds with probability at least $1 - 2e^{-(\sqrt{n}+\sqrt{q})^2}$, where $C_3=C_1+C_2^{-1/2}$ is a positive constant.
Thus, to prove Lemma \ref{comp:lem1}, it suffices to show that, conditional on the event $\mathcal{B}$, the following three inequalities
\begin{align*}
	 & \norm{\widetilde{\z}_k - n^{-1/2}\X\u_k^*}_2
	\lesssim \gamma_{d}^{-1}\widetilde{B}_n,                   \\
	 & \norm{\widetilde{\v}_k - \v_k^*}_2
	\lesssim  \gamma_{d}^{-1}\widetilde{B}_n,                  \\
	 & \frac{\abs{\widetilde{d}_k - \sqrt{n}d_k^*}}{\sqrt{nq}}
	\lesssim \widetilde{B}_n,
\end{align*}
hold uniformly over $k=1, \cdots, r^*$.

It follows from the definition of $\widetilde{B}_n$ that
\begin{align}\label{eq: Bn-tilde-bound}
	\widetilde{B}_n
	= \sigma \left( \frac{1}{\sqrt{n}}+\frac{1}{\sqrt{q}}\right)\leq 2\sigma,
\end{align}
where we use the facts that $q\geq 1$ and $n\geq 1$. Thus, conditional on the event $\mathcal{B}$, we have %$d_1(\E) \leq \sigma C^{\prime}(\sqrt{n}+\sqrt{q})$ and 
\begin{align*}
	\frac{d_1(\E)}{\sqrt{nq}}
	\leq \frac{\sigma C_3(\sqrt{n}+\sqrt{q})}{\sqrt{nq}}
	= C_3\sigma\left( \frac{1}{\sqrt{q}}+\frac{1}{\sqrt{n}}\right)=C_3\widetilde{B}_n\leq c,
\end{align*}
where $c$ is some positive constant. This together with \eqref{eq: B1-Li} and \eqref{eq: B2-Li} yields
\begin{align}
	 & \norm{\widetilde{\z}_k - n^{-1/2}\X\u_k^*}_2
	\lesssim \gamma_d^{-1}\left[1+\frac{d_1(\E)}{\sqrt{nq}}\right]\frac{d_1(\E)}{\sqrt{nq}}
	\lesssim\gamma_{d}^{-1}\widetilde{B}_n, \label{eq: B3-Li} \\
	 & \norm{\widetilde{\v}_k - \v_k^*}_2
	\lesssim \gamma_d^{-1}\left[1+\frac{d_1(\E)}{\sqrt{nq}}\right]\frac{d_1(\E)}{\sqrt{nq}}
	\lesssim\gamma_{d}^{-1}\widetilde{B}_n \label{eq: B4-Li}
\end{align}
for all $k=1, \cdots, r^*$.
An application of Weyl's theorem \citep[Theorem 2]{stewart1998} entails
\begin{align}\label{eq: B5-Li}
	\frac{\abs{\widetilde{d}_k - \sqrt{n}d_k^*}}{\sqrt{nq}}
	=\left|\frac{\widetilde{d}_k}{\sqrt{nq}} - \frac{d_k^*}{\sqrt{q}}\right|
	\leq \frac{d_1(\E)}{\sqrt{nq}}
	\lesssim\widetilde{B}_n
\end{align}
for all $k=1, \cdots, r^*$.
It concludes the proof of Lemma \ref{comp:lem1}.

\bibliographystyle{plainnat}

\end{document}